\documentclass[aps,pre,12pt, reprint]{revtex4-2}
\usepackage[utf8]{inputenc}
\usepackage[T1]{fontenc}
\usepackage{chngcntr}
\usepackage{listings}
\usepackage{upgreek}
\usepackage[version=4]{mhchem}
\usepackage{graphicx,wrapfig,lipsum}
\usepackage{multirow}

\usepackage{amsmath, amsfonts, amssymb}
\usepackage{bm}
\usepackage{xcolor}
\usepackage{siunitx}
\usepackage{ulem}

\usepackage{marginnote}

\newcommand{\bfr}{\bm{r}}

\newcommand{\bfn}{\bm{n}}

\newcommand{\bfv}{\bm{v}}
\newcommand{\bfzeta}{\bm{\zeta}}

\newcommand{\kB}{k_{\mathrm{B}}}
\newcommand{\dd}{\mathrm{d}}
\newcommand{\Tc}{T_\mathrm{c}}

\newcommand{\eg}{\textit{e.g.}}
\newcommand{\ie}{\textit{i.e.}}

\newcommand{\PN}[1]{{\color{cyan} #1}}

\begin{document}

\title{Light-activated Janus particles in geometrically confined  binary solvent }
\author{Micha{\l} Przerwa}
 \affiliation{Institute of Physical Chemistry, Polish Academy of Sciences, Kasprzaka 44/52, PL-01-224 Warsaw, Poland}
\author{Piotr Nowakowski}
\email{Piotr.Nowakowski@irb.hr}
\affiliation{Department of Physical Chemistry, Ru\dj{}er Bo\v{s}kovi\'c Institute, Bijeni\v{c}ka 54, 10000, Zagreb, Croatia}
\affiliation{Max-Planck-Institut f\"ur Intelligente Systeme, Heisenbergstr. 3, D-70569 Stuttgart, Germany}
\author{Takeaki Araki}
\affiliation{Department of Physics, Kyoto University, Kyoto 606-8502, Japan}
\author{Anna Macio\l ek}
\email{amaciolek@ichf.edu.pl}
\affiliation{Institute of Physical Chemistry, Polish Academy of Sciences, Kasprzaka 44/52, PL-01-224 Warsaw, Poland}
\affiliation{Max-Planck-Institut f\"ur Intelligente Systeme, Heisenbergstr. 3, D-70569 Stuttgart, Germany}

\date{\today}

\begin{abstract}
 The coupled dynamics of local fields exert a drastic influence on the light-activated self-propulsion of a  Janus particle in a binary solvent under spatial confinement. In this work, we investigate this problem using numerical simulations that account for local phase separation and wetting phenomena, as well as hydrodynamic effects. We find that confining the binary solvent within a channel results in a reduction of the active particle's propulsion speed and an extension of the duration of its directed motion. Furthermore, the orientational dynamics of this self-propelled particle are not restricted to two dimensions, unlike the phenomenon known as ``orientational quenching''. Increasing the light intensity leads to strong fluctuations in the local fields and, consequently, in the particle's speed. In this context, the significance of key physical parameters governing the efficiency of particle motion control is elucidated.
\end{abstract}
\maketitle

\section{Introduction}
\label{sec:intr}

Active matter, which has been under continuous development for over two decades, has become an important branch of the condensed matter field. Progress in this field is so diverse and intensive that it has led to the publication of a metareview, a survey of active matter reviews summarizing the impressive body of existing research~\cite{te2025metareview}. Many fundamental questions concerning both individual active objects and their ensembles have already been answered, hence the main research stream has shifted toward more complex collective behavior~\cite{RevModPhys.97.015007,zaferani2026sperm}, interactions of active systems with their environment~\cite{granek2024colloquium}, optimal control in  active matter~\cite{fbgp-qpvv,davis2024active,alvarado2026optimal} or increasingly sophisticated active systems, \eg, pulsating~\cite{PhysRevLett.131.238302}, chiral~\cite{Liebchen_2022} or intelligent~\cite{D2SM00736C,bauerle2018self} active matter, among others~\cite{Gompper_2025}.

Considerable insights into behavior of natural  or man-made  micro- and nano-sized objects capable of self-propulsion (also referred to as microswimmers) is  gained based  on minimal models of active particles, such as active Brownian particles~\cite{PhysRevLett.99.048102,MARCHETTI201634,romanczuk2012active}, run-and-tumble particles~\cite{solon2015active,schnitzer1993theory}, and active Ornstein-Uhlenbeck particles~\cite{haunggi1994colored,PhysRevE.90.012111,sepulveda2013collective,PhysRevE.91.042310,PhysRevE.103.032607}. In these models, all interactions with the solvent, including hydrodynamic ones, are ignored and it is assumed that the solvent only provides friction, which causes overdamped dynamics.  The success of these models lies in the fact that although the different types of active particles differ significantly in detail, they share the common characteristic of being persistent random walkers. This means that compared to the passive random walker, a new scale emerges separating ballistic motion from diffusive behavior, which occurs on a larger scale. This new scale is determined by the average distance that the active particle travels in approximately one direction, called the persistence length, and the time it takes to travel it. The minimal models of active particle  are  popular for  addressing  fundamental questions about the nonequilibrium statistical mechanics of active systems.

A different approach to the description of active systems is necessary when solvent-mediated hydrodynamic interactions are important.  In such a case, the solvent dynamics should be included in the model~\cite{Elgeti_2015,RevModPhys.85.1143,Zottl_2016}.  The standard model focusing solely on hydrodynamic flow generated by self-propelled objects in  the limit where viscous forces dominate over inertial effects (small Reynolds number) is the so-called squirmer~\cite{lighthill1952squirming,blake1971spherical,10.1063/5.0300392,Lauga_2009}. Squirmer hydrodynamics relies on prescribed surface slip velocities, without taking the underlying mechanism into account, and the flow fields induced by these slip velocities. In the far-field limit, the squirmer generates a flow that can be systematically decomposed into a hierarchy of Stokes flow singularities, with the dominant term being the stresslet, decaying with the distance as $r^{-2}$. 
This reflects the dipole nature of the squirmer's propulsion and the sign of dipole  determines  its ``pusher'' or ``puller'' character. In the presence of obstacles, external walls or other self-propelled objects, far-field approximations (force or source dipoles) break down and one has to consider near-field hydrodynamics of squirmers~\cite{yoshinaga2017hydrodynamic,PhysRevLett.112.118101}.

In the case of artificial self-propelled objects, such as active colloids, there is another aspect that has to be taken into account when considering the interactions of active particles with boundary surfaces, complex environments, or other active particles. In such situations, the specific forms of the local thermodynamic fields created by the particles in order to self propel,  play a crucial role in the dynamic behavior~\cite{Zottl_2016}. Examples of  self-generated local thermodynamic fields are the temperature field~\cite{PhysRevLett.105.268302,truong2026light}, the chemical concentration field~\cite{golestanian2007designing,popescu2016self}, the composition field~\cite{Volpe2011,Buttinoni_2012}  or the electric field~\cite{paxton2004catalytic};  the mechanism of self-locomotion caused by gradients of these fields is called self-phoresis in the literature. The Anderson theory of phoresis~\cite{anderson1989colloid} employed to self-generated, rather than externally imposed, thermodynamic fields gradient, predicts the surface slip velocities proportional to the field gradient along the surface~\cite{popescu2016self,wurger2015self}.  This can be used as input for squirmer hydrodynamics as long as the surface velocity field is stationary. When the active colloid is close to other objects or walls, this may not be the case, and then the coupled evolution equations for the thermodynamic field and solvent flow must be solved in a self-consistent manner under appropriate boundary conditions  at the colloid surface~\cite{de2013phoretic,yang2014hydrodynamic,popescu2018effective}. 
Of course, this is quite a difficult task and can only be accomplished using numerical methods. In recent years, a relatively large number of works have been published implementing this program for catalytically active particles~\cite{uspal2015rheotaxis,mozaffari2016self,popescu2009confinement,uspal2015self}. In contrast, for light-activated colloids, to the best of our knowledge, only three such publications exist and they deal exclusively with self-propulsion in   bulk~\cite{samin2015self,gomez2017tuning,Araki2019}. 

Light-activated colloids, such as Janus particles coated by light-absorbing material,  self-propel under  permanent  illumination by laser light~\cite{PhysRevLett.105.268302,Bregulla:13,Buttinoni_2012,Volpe2011}. They represent attractive  alternatives to the fuel-dependent  catalytic active particles and  offer significant advantages in microscale manipulation thanks to high spatiotemporal precision, remote controllability, and non-invasive activation provided  by light. For Janus particles suspended in a binary mixture with a lower critical point of demixing, illumination with low-intensity laser light induces two local thermodynamic gradients: a temperature gradient resulting from local heating and a compositional gradient resulting from local phase separation.  To account for these local fields, in Refs~\cite{samin2015self,gomez2017tuning} the coupled equations of the evolution of the composition field and its flow around the colloid were solved in the stationary temperature profile obtained from the Laplace equation. In the steady state, the self-propulsion velocity was determined based on the requirement that the colloid should be force-free. 
An even more complete description is provided in Ref.~\cite{Araki2019}, where the full evolution equations for all three relevant thermodynamic fields are solved simultaneously and in a self-consistent manner using the fluid particle dynamics (FPD) method. Furthermore, in this paper, the wetting properties of the Janus particle are treated more appropriately than in~\cite{samin2015self,gomez2017tuning}: they arise from surface interactions with the composition field, which is consistent with the field-theoretic description of a binary mixture, rather than being dictated by the contact angle as in macroscopic theories. The advantage of the FPD method is that it does not require defining a boundary condition for the velocity at the colloid surface and that the self-propelling velocity can be determined directly.
The results of both approaches could explain some of the experimental findings and show that  self-propulsion in this case is not achieved by slip velocities on the particle surfaces, \ie, it is not the result of self-diffusiophoresis, as proposed in  Ref.~\cite{wurger2015self}. 

As mentioned above, the calculations presented in references~\cite{samin2015self,gomez2017tuning,Araki2019} concern the motion of an active particle in a bulk system. However, typical experiments utilizing light-activated Janus particles are conducted in confined geometries. In such cases, both the temperature field and the concentration field surrounding the particle are perturbed by the presence of bounding surfaces~\cite{C8SM01258J,D0SM00964DF,lozano2016phototaxis}. Furthermore, the concentration field typically interacts with the walls in a manner dependent on their wetting properties. The flow induced by the Janus particle is also modified as a result of the introduction of geometric constraints. Moreover, self-propelled particles may be subject to hydrodynamic self-interactions; this phenomenon occurs when the flow field generated by a single active particle is influenced by the wall surfaces and subsequently exerts a feedback effect on that same active particle~\cite{Liebchen_2022}. All of this, of course, influences the self-propulsion that sets the light-activated Janus particles in motion, as well as the motion itself. Until now, these aspects had not been investigated in the literature, which provided the motivation for this  study. 

It is worth noting here that the problem of self-propulsion in confined geometries has been studied quite extensively within the framework of the simplified models of natural or artificial microswimmers discussed above. This stems from the significance of this issue both for understanding biological processes, such as the motility of sperm within the female reproductive tract~\cite{denissenko2012human}, and for engineering applications, including, for instance, microfluidic bacterial separation or targeted drug delivery systems~\cite{elgeti2016microswimmers,Bechinger2016}. Previous studies have focused on boundary conditions ranging from simple flat walls to complex, tight, and curved geometries.
Key aspects investigated include: attraction and accumulation induced by boundaries, changes in swimming speed, the phenomenon of trapping~\cite{PhysRevLett.112.118101,10.3389/fphy.2022.926609}, alignment of orientation~\cite{10.1063/1.4981886}, and the emergence of collective, highly organized motility patterns. Research results indicate that active particles tend to accumulate near boundaries, a phenomenon resulting from hydrodynamic and steric (physical) interactions~\cite{elgeti2013wall}. A ``gliding'' type of motion along the walls is frequently observed among them. Detailed analyses classify particle behaviors based on various swimming modes---such as ``gliding,'' ``trapping,'' ``spinning,'' or ``bouncing''---depending on whether a given particle is a ``pusher'' (\eg, \textit{E. coli}) or a ``puller'' (\eg, \textit{Chlamydomonas})~\cite{spagnolie2012hydrodynamics,berke2008hydrodynamic}. In circular tubes or narrow channels, microswimmers often adopt helical trajectories or oscillate between the walls~\cite{PhysRevFluids.9.083302,lauga2006swimming}.  Regarding self-propulsion speed, various studies demonstrate that within channels, it undergoes significant changes due to hydrodynamic interactions with the boundaries, spatial confinement effects, and altered rotational dynamics. Although in certain scenarios spatial confinement may lead to increased drag and reduced speed~\cite{zhu2013low,C5SM01412C}, it often results in increased swimming speed or persistent, directed motion (trapping) compared to behavior in an unbounded medium~\cite{ledesma2013enhanced,liu2016bimetallic,popescu2009confinement,das2015boundaries}.

In this work, we consider a simple slit channel, a geometry that arises naturally in the experiments described in references~\cite{Volpe2011,Buttinoni_2012,gomez2017tuning,bauerle2018self,lozano2016phototaxis,PhysRevLett.116.138301,PhysRevLett.121.078003}. Our objective is to investigate the extent to which this confining geometry influences the motility of light-activated Janus colloids, particularly in the case of narrow channels. Furthermore, we aim to deepen the analysis of the applicability of the FPD method~\cite{Tanaka2000,furukawa2018physical} for describing interacting active colloids, which constitutes a first step toward addressing pairwise interactions and many-body problems. In the case of passive colloids, the FPD method has proven effective in accurately simulating the dynamics  while simultaneously accounting for full hydrodynamic interactions, inertial effects, incompressibility, thermal noise, and additional fluid degrees of freedom. 

The paper is organized as follows: In Sec.~\ref{sec:1} we introduce the system and  provide details of the FPD approach adopted to our problem.  Section~\ref{sec:2} describes  the numerical setup. In Sec.~\ref{sec:res} we report our results and in Sec.~\ref{sec:dis} we discussed them and make connection with existing experimental findings. 

\section{Method}
\label{sec:1}

\begin{figure}[t]
        \centering
         \includegraphics[width=0.99\columnwidth]{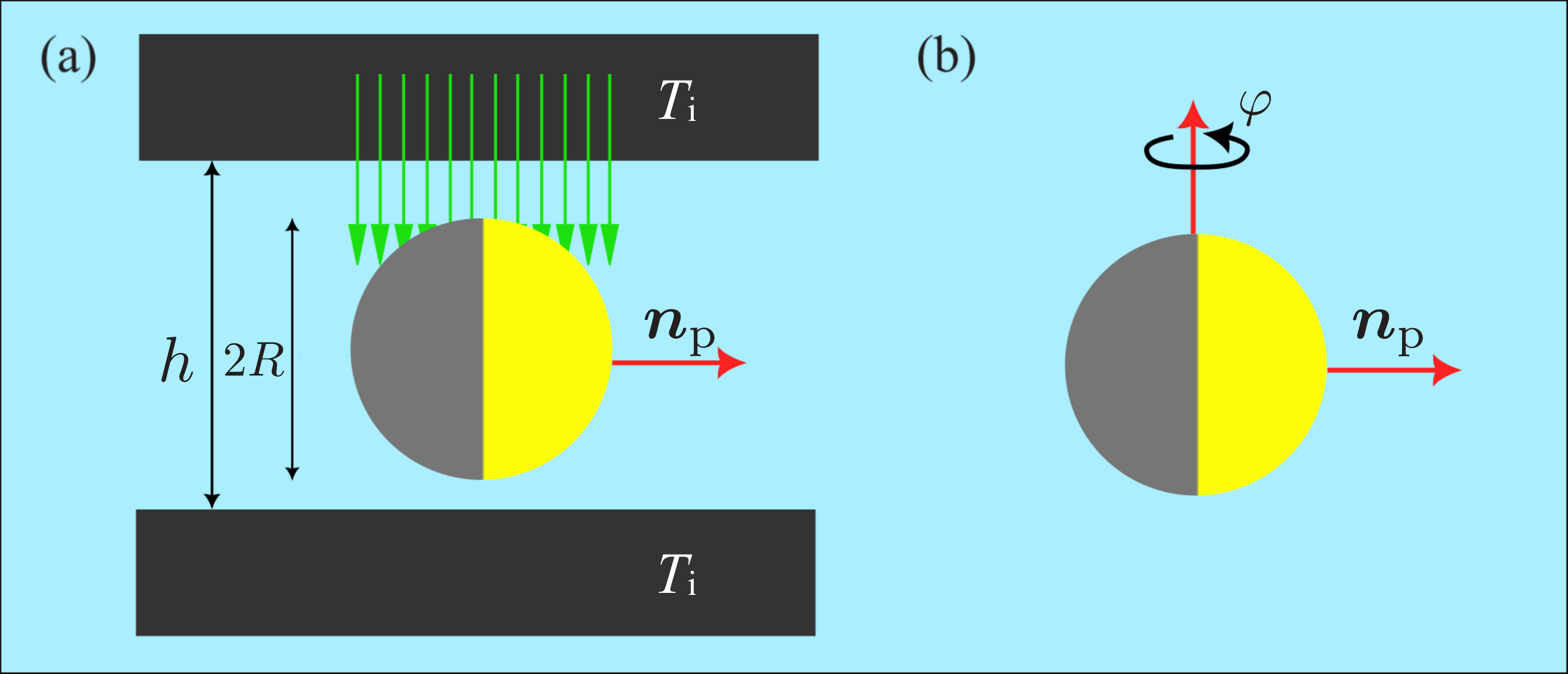}
    \caption{Schematic plot of (a) the system and (b) the Janus particle. Two parallel walls enclosing the system are separated by a distance $h$ and kept at constant temperature $T_\mathrm{i}$. The particle has a shape of a sphere of radius $R$ with the location of light-heated hemisphere defined by a unit vector $\bfn_\mathrm{p}$. We use the angle $\varphi$ to describe the rotations of particle around horizontal axis.}

        \label{fig:1}
        \end{figure}

We consider a spherical colloidal particle of radius $R$ suspended in a mixture of two liquids with a lower critical point (LCP) of demixing. The system is enclosed between two parallel walls separated by the distance $h$ that are kept at a fixed temperature $T_\mathrm{i}$, slightly below the lower critical temperature $\Tc$ of the mixture. This setup ensures that, away from the colloid, the binary mixture is in a mixed state. Schematic plot of our system is presented in Fig.~\ref{fig:1}(a).

The colloid immersed in the mixture is assumed to be Janus particle, \ie, two hemispheres of its surface have different adsorption properties and they accumulate different components of the mixture around them. Moreover, inspired by the pioneering experiments \cite{Buttinoni_2012,Volpe2011}, we assume that one of the hemispheres (so-called ``cap''), defined by the orientation vector $\bfn_{\rm p}$, can absorb light and, when illuminated by an external laser, it heats up. Such a heating leads to a local increase of temperature of the mixture and can cause local phase separation. As observed in the experiments, this phenomenon creates active propulsion of the colloidal particle and produces its directed motion. We note that our numerical routines allow for more complicated shapes of patch on the colloid. Analysis of these cases, however, goes beyond the scope of the manuscript. 

We describe the configuration of the Janus particle with the position vector $\bfr_{\mathrm p}$, pointing to the center of the particle, and the orientation vector $\bfn_{\rm p}$ - a unit vector pointing from the center of the particle in the direction of the center of the light -absorbing hemispherical cap.

To study our system we use the fluid particle dynamics method---theoretical framework and its numerical implementation developed by Tanaka and Araki~\cite{Tanaka2000} for the description of colloidal suspensions---as it is capable of capturing the main features of our non-equilibrium system, in particular the self-propulsion of the Janus particle~\cite{Araki2019}.
Within this approach, the Janus particle is described using two functions: The shape function
\begin{subequations}
\begin{equation}\label{eq:particle}
\psi\left(\bfr; \bfr_{\mathrm{p}}\right) = \frac{1}{2}\left[1+\tanh\left(\frac{R-\left|\bfr-\bfr_{\mathrm p}\right|}{d_\psi}\right)\right],
\end{equation}
is equal to 1 when $\bfr$ is inside the colloid and 0 if it is outside, with a smooth crossover on the surface of sphere on the lengthscale $d_\psi$. The orientation function
\begin{equation}\label{eq:orientation}
\theta\left(\bfr; \bfr_{\mathrm{p}}, \bfn_{\rm p}\right) =\frac{1}{2}\left[1+\tanh\left(\frac{1}{d_\theta} \bfn_{\rm p}\cdot\frac{\bfr-\bfr_{\mathrm p}}{\left|\bfr-\bfr_{\mathrm p}\right|}\right)\right],
\end{equation}
\end{subequations}
is equal 1 for all points $\bfr$ facing the light-absorbing hemisphere and 0 for points facing the other hemisphere, with a smooth crossover on the angular lengthscale $d_\theta$. 

We note that the FPD method is based on a hybrid approach, which combines a lattice simulation for continuous fields and an off-lattice simulation for particles. In the numerical implementation, we use $\bfr$ for an on-lattice site to describe the fields, and the position $\bfr_{\mathrm p}$ of the particle can be an arbitrary vector.

In order to describe our nonisothermal system we use coarse-grained entropy and energy, from which temperature is defined as a functional of the order parameter and energy density. This approach has been developed by Onuki~\cite{Onuki2005,Onuki2007} for a one-component mixture and extended to a binary liquid mixture in Ref.~\citenum{Gonnella2008}. To describe the mixture we use order parameter $\phi=(\rho_1-\rho_2)/(\rho_1+\rho_2)$, where $\rho_1$ and $\rho_2$ denote number densities of the two components of the binary mixture. We assume that the mixture is symmetric, \ie, $\bar{\phi}=0$ in the mixed state.

The local  energy per molecule is a function of local composition $\phi\left(\bfr\right)$ and  temperature $T\left(\bfr\right)$, as well as of the position $\bfr_{\mathrm{p}}$ and the orientation $\bfn_{\rm p}$ of the particle:

\begin{multline}
\label{eq:e_phi}
e\left(\phi,T;\bfr_{\rm p},\bfn_{\rm p}\right)=\left(1-\psi\right)\left[-\frac{\varepsilon}{2}\phi^2+\frac{c}{2}\left(\nabla\phi\right)^2\right]\\
+\frac{\chi_{\mathrm{p}}}{2}\left(\phi-\phi_{\mathrm{p}}\right)^2\psi +\left[W_{\mathrm{u}} \left(1-\theta\right)+W_{\mathrm{c}}\theta\right]\left|d_\psi\nabla\psi\right|\phi\\
+\frac{3}{2}(1-\psi)k_{\rm B}T.
\end{multline}
 Here and in the following $\nabla$ denotes differential operator with respect to $\bfr$.
The first term on the right-hand side of Eq.~\eqref{eq:e_phi} represents the internal energy of the solvent. Since we are interested in the phase region near the critical point, we employ the Ginzburg-Landau form here, featuring an interaction parameter $\varepsilon$ and a gradient coefficient $c$ related to the range of interaction of particles, which we both assume to be independent of temperature. 
 The prefactor $(1-\psi)$ ensures that this energy is calculated only outside the colloid. Second term is a contribution that prevents solvent invasion into the particle with two positive control parameters $\chi_{\mathrm{p}}$ and $\phi_{\mathrm{p}}$  (see Ref.~\citenum{Tanaka2000}). The third term in Eq.~\eqref{eq:e_phi} is the surface contribution, 
  describing the interaction of the solvent with the surface of the Janus particle with different wetting parameters $W_{\mathrm{c}}$ and $W_{\mathrm{u}}$ for capped and uncapped hemisphere, respectively. The factor $\left| d_\psi\nabla\psi \right|$ makes this contribution nonzero only in the vicinity of the surface of the spherical colloid.
 Finally, the last term in Eq.~\eqref{eq:e_phi} is the kinetic energy per molecule with the Boltzmann constant $k_{\rm B}$.

For the entropy per molecule close to the critical point of demixing, we take the expansion around $\phi=0$  up to the fourth order
\begin{multline}\label{eq:entropy}s\left(\phi,T;\bfr_{\rm p}\right)=\\
\kB\left(1-\psi\right)\left[-\left(\frac{a}{2}\phi^2+\frac{b}{4}\phi^4 \right)+\ln\left(\frac{d_0^3}{\lambda_{\mathrm{T}}^3}\right)+1\right],
\end{multline}
where $a$ and $b$ are positive constants, $\lambda_{\mathrm{T}}$ (proportional to $T^{-1/2}$) is the thermal de Broglie length, and $d_0$ denotes the molecular size, which is assumed to be equal to $v_0^{1/3}$ and constant, where $v_0$ is the inverse  of the number density of a liquid mixture.

The total energy and entropy are given by 
\begin{equation}\label{eq:total_e_s}
 \mathcal{E}=\frac{1}{d_0^3}\int \dd\bm{r}\,e, \qquad  \mathcal{S}=\frac{1}{d_0^3}\int \dd\bm{r}\,s.
\end{equation}
From energy and entropy, we define the local chemical potential difference as \footnote{We note that, the exact formula for the chemical potential depends on the integration measure used in the functional $\mathcal{S}$. Here, following \cite{Gonnella2008}, we assume that the integral is done using dimensionless variable $\bfr/d_0$. For other choices, additional prefactors must be included in Eq.~\eqref{eq:chempot}.} 
\begin{equation}
\label{eq:chempot}
 \mu = -T \left(\frac{\delta}{\delta \phi}\mathcal{S}\right)_{e}.
\end{equation}
After some algebra we get
\begin{multline}
\label{eq:mu}
\frac{\mu}{\kB T}=\left(1-\psi\right)\left[\left(a-\frac{\varepsilon}{k_{\rm B}T}\right)\phi+ b\phi^3\right]\\
-\frac{c}{k_{\rm B}T}\nabla\cdot \left[\left(1-\psi\right)\nabla\phi\right]\\
+\frac{1}{\kB T}\left\{\left[W_{\mathrm{u}}\left(1-\theta\right)+W_{\mathrm{c}}\theta\right]\left|d_\psi\nabla \psi\right|+\chi_{\mathrm{p}}\left(\phi-\phi_{\mathrm{p}}\right)\psi \right\}.
\end{multline}
To facilitate comparison with experiments, we focus on the universal features of the system near the LCP, which are the same as for the upper critical point. In binary liquid mixtures, the LCP arises from specific molecular interactions, such as hydrogen bonding.
Rather than considering the detailed temperature dependence of $\varepsilon$, we replace the quantity ${a-\varepsilon/(k_{\rm B}T)}$ in Eq.~(\ref{eq:mu}) by $-\tau$ and assume a simple relation to $T$, namely $\tau=(T-\Tc)/|T_{\rm i}-\Tc|$, where $\Tc=\varepsilon/(ak_{\rm B})$.
In our simulations we study the time evolution of the field $\tau\left(\bfr\right)$ rather than the temperature field $T\left(\bfr\right)$. With our definition of $\tau$, the symmetric mixture undergoes phase separation for $\tau>0$ ($T>T_{\rm c}$). 
For simplicity, in our simulations the factors $1/(k_{\rm B}T)$ in the second and third terms on the right-hand side of Eq.~(\ref{eq:mu}) were replaced by $1/(k_{\rm B}\Tc)$. 
In an isothermal homogeneous system ($\tau<0$), mean-field theory predicts the bulk correlation length to be $\xi=\sqrt{c/(k_{\rm B}\Tc|\tau|)}$. 
In the present study, the correlation length at the initial temperature $T_{\rm i}$ is set to $2d_0$.

We assume that time evolution of the concentration field $\phi$, temperature field $\tau$, and flow field $\bm{v}$ is given by the following dynamic equations~\cite{Araki2019, GomezSolano2020a, *GomezSolano2020b}
\begin{subequations}\label{eq:devs}
\begin{align}
\label{eq:dev_phi}\frac{\partial \phi}{\partial t}&=-\nabla \cdot \left(\phi \bfv\right)-\nabla \cdot \left[-L_0(1-\psi) \nabla \frac{\mu}{\kB T}\right] + \nabla \cdot \bfzeta,\\
\label{eq:dev_T}  
\frac{\partial \tau}{\partial t}&=-\nabla \cdot \left(\tau \bfv\right)+\nabla \cdot \left(L_\mathrm{T}\nabla \tau\right)+H\theta\left|d_\psi\nabla \psi\right|,
\\
\label{eq:dev_v}
\frac{\partial \rho \bm{v}}{\partial t}&=\bm{f}-\nabla p-\nabla\cdot\mathsf{\Pi}+\nabla\cdot\mathsf{\sigma}, 
\end{align}
\end{subequations}
where ${\bfzeta}$ is the thermal noise, which we assume to be white Gaussian noise with  an amplitude $\sqrt{\langle |\bfzeta(\bfr,t)|^2\rangle}=\zeta_0$ imposed at each site of the lattice.  $L_0$ and $L_{\rm T}$ are positive kinetic coefficients, which we assume to be independent of  $\phi$ and $T$. 
For the isothermal system, $L_0$  is related to the  mixture inter-diffusion coefficient $D_0$~\cite{binder2007interdiffusion,10.1063/1.1568333}. We note that, although the mixture inter-diffusion coefficient varies quite a bit with temperature and vanishes at the critical temperature due to the divergence of concentration fluctuations (critical slowing down), the Onsager coefficient $L_0$ remains finite or diverges much more weakly, representing the ``bare'' mobility of the molecules. For aqueous mixtures, such as water–2,6-lutidine, in the temperature range that we observe around the colloid\PN{,} the inter-diffusion coefficient  $D_0$ is of the order $10^{-12}$--$10^{-11}\,$\unit{\square\meter/\second}, except of the critical temperature~\cite{10.1063/1.2188396}, whereas $L_0 \simeq \qty{3.0e-11}{\square\meter/\second}$. 
Here  we  neglect the off-diagonal kinetic coefficients, which cause the so-called Soret and Dufour effects.  

In the temperature diffusion equation (\ref{eq:dev_T}), 
we distinguish between the thermal diffusion speed within the particle by assuming $L_\mathrm{T}=L_{\mathrm{T0}}\left(1-\psi\right)+L_{\mathrm{Tp}}\psi$, \ie, different thermal diffusion constants $L_{\mathrm{T0}}$  and $L_{\mathrm{Tp}}$ of the binary solvent and within the particle, respectively.
The last term represents the power of the heat source, which is located in the cap of the particle represented  by $\theta\left|d_\psi\nabla\psi\right|$.
The coefficient $H$ is given by: 
\begin{equation}\label{eq:cool_pow}
H=g\left\{\tau_{\mathrm{s}}-\tau\left(\bfr,t\right) \right\},  
\end{equation}
where $\tau_{\rm s}$ is the target temperature of the particle capped surface. 
The first term, $g\tau_{\rm s}$, is the heating power divided by the heat capacity, and the second one, $-g\tau$, represents the dissipation to the bath.  In such an approach,  dissipation is introduced only at the surface. This, together with fixed temperature at the walls of the simulation box, controls the location of the critical isotherm around the colloid. Accounting for dissipation at the colloid surface significantly reduces the simulation time compared to the case where $H$ is simply a constant.
For comparison, we present in the Appendix~\ref{app:A} the results obtained with a constant value of  heating power.

In the hydrodynamic equation (\ref{eq:dev_v}), $\rho$ is a mass density; we assume that the density of the liquid is the same as that of the particle.  The first term of the right hand side represents the force   field stemming from the motion and rotation of the particle, as well as its interaction with the walls
\begin{multline}
\bm{f}=-\frac{\mathcal{\psi}}{\Omega}
\left(\frac{\partial \mathcal{E}}{\partial {\bfr_{\textrm p}}}\right)_s
-\frac{1}{2}\nabla \times 
\left[\frac{\psi}{\Omega}{\bfn}_{\rm p}\times \left(\frac{\partial \mathcal{E}}{\partial \bfn_{\rm p}}\right)_s\right]\\
+\frac{\psi}{\Omega}\bm{F}_\text{walls},
\end{multline}
where $p$ is the pressure, which is introduced to satisfy the condition of incompressibility of the fluid $\nabla\cdot \bm{v}=0$. The force $\bm{F}_\text{walls}$ acting on the Janus particle stems from its interaction with the walls, see the discussion below.
The mechanical stress $\mathsf{\Pi}$, which stems from the interface tension, and viscous stress $\mathsf{\sigma}$ tensors are given by
\begin{subequations}
\begin{eqnarray}
    \label{eq:mstress}\mathsf{\Pi}&=&\frac{c}{d_0^3}\nabla \phi:\nabla \phi,\\
    \label{eq:vstress}\mathsf{\sigma}&=&\left[\eta\left(1-\psi\right)+\eta_{\rm p}\psi\right]\left[\nabla:\bm{v}+(\nabla:\bm{v})^{\rm T}\right], 
\end{eqnarray}
\end{subequations}
in which $\eta$ and $\eta_{\mathrm{p}}$ are the viscosities of the solvent and within the particle, respectively. 

The particle position and orientation are transported by the flow and its vorticity according to 
\begin{subequations}\label{eq:particle_evolution}
\begin{equation}
\bm{v}_{\rm p} =\frac{1}{\Omega}\int \dd\bfr\, \psi\left(\bfr,\bfr_{\rm p}\right)\bfv\left(\bfr\right),
\end{equation}
and 
\begin{equation}
\frac{\dd}{\dd t}\bfn_{\rm p}=\frac{1}{2\Omega}\int \dd\bfr \left[\psi\left(\bfr;\bfr_{\rm p}\right)\bfn_{\rm p}\times \left(\nabla\times \bfv\left(\bfr\right)\right)\right],
\end{equation}
\end{subequations}
where $\Omega=\int \dd\bfr\,\psi\left(\bfr;\bfr_{\rm p}\right)$ is the volume of the particles. 
$\bm{v}_{\rm p}(={\rm d}\bfr_{\rm p}/{\rm d}t)$ represents the velocity of the particle.

Finally, we model Janus particle--walls interaction by a Weeks--Chandler--Anderson potential
\begin{equation}\label{eq:wall_potential}
    U_\text{wall}\left(r_\mathrm{w}\right) = 
    \begin{cases}
        4\epsilon_{\rm w}\left[\left(\sigma_{\rm w}/r_\mathrm{w}\right)^{12}-\left(\sigma_{\rm w}/r_\mathrm{w}\right)^{6}\right] + \epsilon_\mathrm{w}\hspace*{-1.7cm} & \\
        &\text{for }r_\mathrm{w}\leqslant 2^{1/6}\sigma_{\rm w},\\
        0 & \text{for }r_\mathrm{w}>2^{1/6}\sigma_{\rm w},
    \end{cases}
\end{equation}
where $r_\mathrm{w}$ is the distance between the center of the particle and the wall. This is the Lennard--Jones potential, shifted upright by $\epsilon_{\rm w}$ and truncated at its minimum at $r_\mathrm{w}=2^{1/6}\sigma_{\rm w}$ ~\cite{Heyes2006}. In the following we assume $\epsilon_{\rm w}=(1/4)k_{\rm B}\Tc$ and $\sigma_{\rm w}=R$, the radius of the particle. 
We use two potentials of the form given in Eq.~\eqref{eq:wall_potential} to calculate the force $\bm{F}_\text{walls}$
acting on the particle from both walls, included in Eq.~\eqref{eq:dev_v}.
We assume that both top and bottom walls of the channel prefer the same component of the binary solvent with the same strength. In our model this is accounted for through  surface contribution to the chemical potential---specifically, $\left. \mu /(k_BT)\right|_{\rm wall}=h_{\mathrm w}$, where $h_{\mathrm w}$ is a surface field that depends on the details of the actual interaction between channel  surfaces and the two species of the solvent (also referred to as the wetting parameter). This contribution stems from the wetting energy $e_{\mathrm{wall}}(\phi,T)=h_{\mathrm w}\phi$ at the wall surfaces.

\section{Numerical setup}
\label{sec:2}

We study our system numerically on a three dimensional lattice with lattice spacing equal to the molecular size $d_0$. We assume that there are periodic boundary conditions in $x$ and $y$ directions and the walls limit our system in $z$ direction. We solve the equations of motion~\eqref{eq:devs} using the Marker-and-Cell method~\cite{harlow1965numerical} with a staggered lattice.

In the calculations we use $d_0$ as a unit of length, characteristic time of concentration diffusion $t_0=d_0^2/L_0$ as a unit of time, and $\kB\Tc$ as a unit of energy. The size of the simulation box is $256\,d_0$ in $x$ and $y$ directions, and the walls are placed at $z=0$ and $z=h$ with various values of $h$ between $32\,d_0$ and $128\,d_0$. We assume  non-slip boundary conditions for the hydrodynamic flow $\bfv\left(z=0\right)=\bfv\left(z=h\right)=\bm{0}$, and impose vanishing concentration flux $-L_0\nabla \left(\mu/\kB T\right)=\bm{0}$. 
10
We choose an initial temperature corresponding to the mixed phase of a binary solvent $T_{\rm i}=0.98\Tc$, which corresponds to $\tau=-1$.
After turning the heating on, we control the temperature on the cap by changing $g$ with $\tau_{\rm s}=6.631$. 

\begin{figure*}[t]
 \centering
    \includegraphics[width=\textwidth]{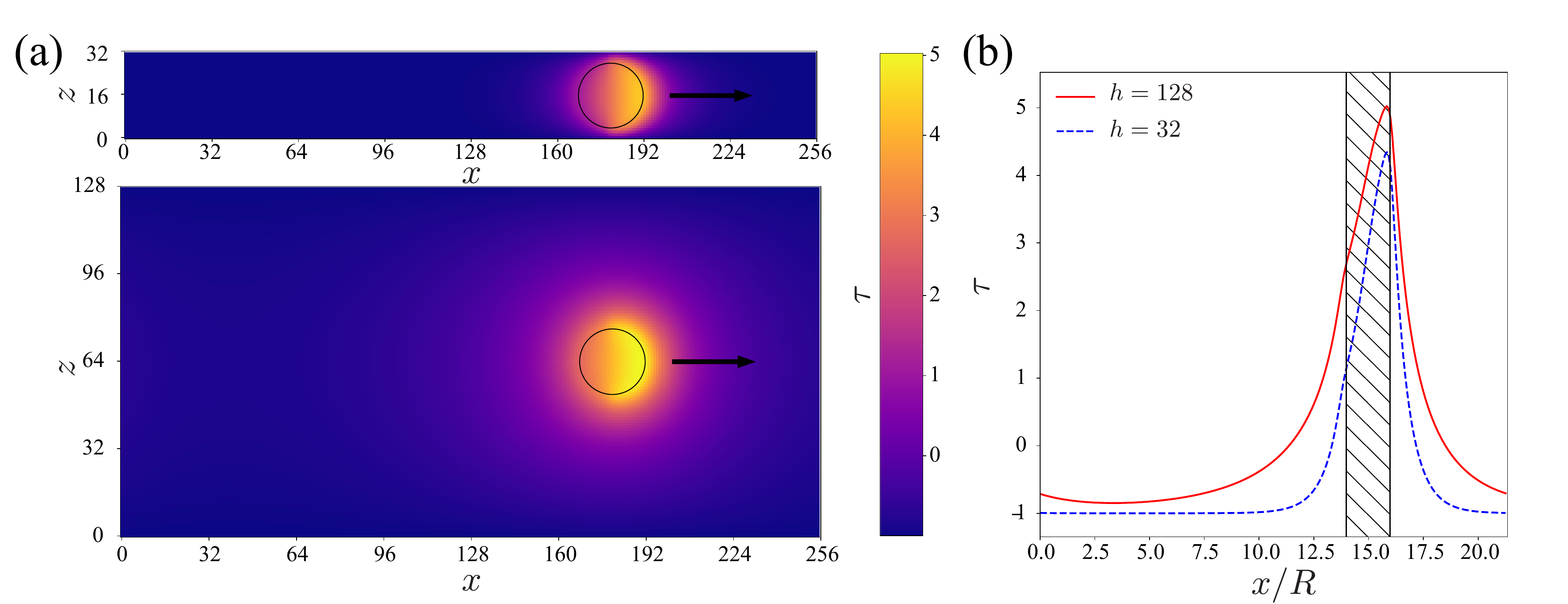}
       \caption{ (a) A snapshot of the temperature field around a Janus particle at time $t=100$ after switching on the illumination for two channels with widths $h=32$ (top) and $h=128$ (bottom). The wall temperature is $\tau=-1$ and the heating power coefficient $g=50$. The black circle represents the particle; its initial orientation along $x$-axis ($\varphi_{\mathrm{i}}=0$) remains practically unchanged, as indicated with black arrow.
       (b) Corresponding temperature fields profiles along the $x$-axis  passing through the center of the particle. The hatched region indicates the interior of particle. To facilitate the comparison between two channel widths, we have use the periodic boundary conditions in $x$ direction to align the particles in the same point of the channel. The same shift has been done in Figs.~\ref{fig:3} and~\ref{fig:4}.} 
\label{fig:2}
\end{figure*}

By dividing the hydrodynamic equation (\ref{eq:dev_v}) by $\rho d_0/t_0^2$, it is expressed in a dimensionless form. 
The coefficients of the third and fourth terms in 
the dimensionless form of Eq.~(\ref{eq:dev_v}) are given by $c/(\rho d_0^3 L_0^2)$ and $\eta/(\rho L_0)$, respectively.
Since inertial effects are negligible in the system considered here, we chose the parameters $c/(\rho d_0^3 L_0^2) = 400$, which corresponds to $\rho/(k_{\rm B}T_{\rm c}/d_0L_0^2)=0.01$, 
and $\eta/(\rho L_0) = 50$ with $\eta_{\rm p}/\eta=50$ to suppress inertia further. 
Under these conditions, we integrate the hydrodynamic equation \eqref{eq:dev_v} with the time step $10^{-5}t_0$. The fields $\phi$ and $\tau$, as well as the position $\bfr_\mathrm{p}$ and orientation $\bfn_\mathrm{p}$ of particle are updated using Eqs.~\eqref{eq:dev_phi}, \eqref{eq:dev_T}, and \eqref{eq:particle_evolution} only once per 100 iterations of hydrodynamic equation, making the overall time step of our model $\Delta t=10^{-3}t_0$.

We assume that the capped surface is strongly hydrophilic with $W_{\mathrm{c}} = -8\,\kB\Tc$, while the uncapped surface is neutral  $W_{\mathrm{u}}= 0$. The walls are assumed to be weakly hydrophilic with $h_{\mathrm{w}}=-0.4\,\kB\Tc$ or weakly hydrophobic with $h_{\mathrm{w}}=0.4\,\kB\Tc$. In Sec.~\ref{subsec:res3}, we also consider $h_{\mathrm w}=\pm 4\,\kB\Tc$ and $\pm 8\,\kB\Tc$. 

For the convenience of the reader, all model parameters used in the numerical calculations are listed in the Appendix~\ref{app:B}. For the sake of simplicity, in the section below all numbers are given in the simulation units.

\section{Results}
\label{sec:res}
\subsection{Active particle in the channel}
\label{subsec:res1}

 We first investigate how confinement affects active motion. For this purpose, we placed a Janus particle of radius $R=12$ in a symmetric mixture with an average concentration of $\bar{\phi}=0$, confined between two flat walls. The initial position of the Janus particle is in the center of a channel with orientation $\bfn_{\rm p}=(\cos \varphi_{\rm i},\sin\varphi_{\rm i},0)$ in the $xy$-plane.  At $t=0$, we turn on the illumination of  the system by imposing $g>0$.  After a sufficiently long heating (see Sec.~\ref{subsec:2} for more details), the system reaches a steady state, in which  the particle moves along $\bm{n}_{\rm p}$ with  its heated part in front with a constant velocity. 
\begin{figure*}[t]
 \centering
       \includegraphics[width=\textwidth]{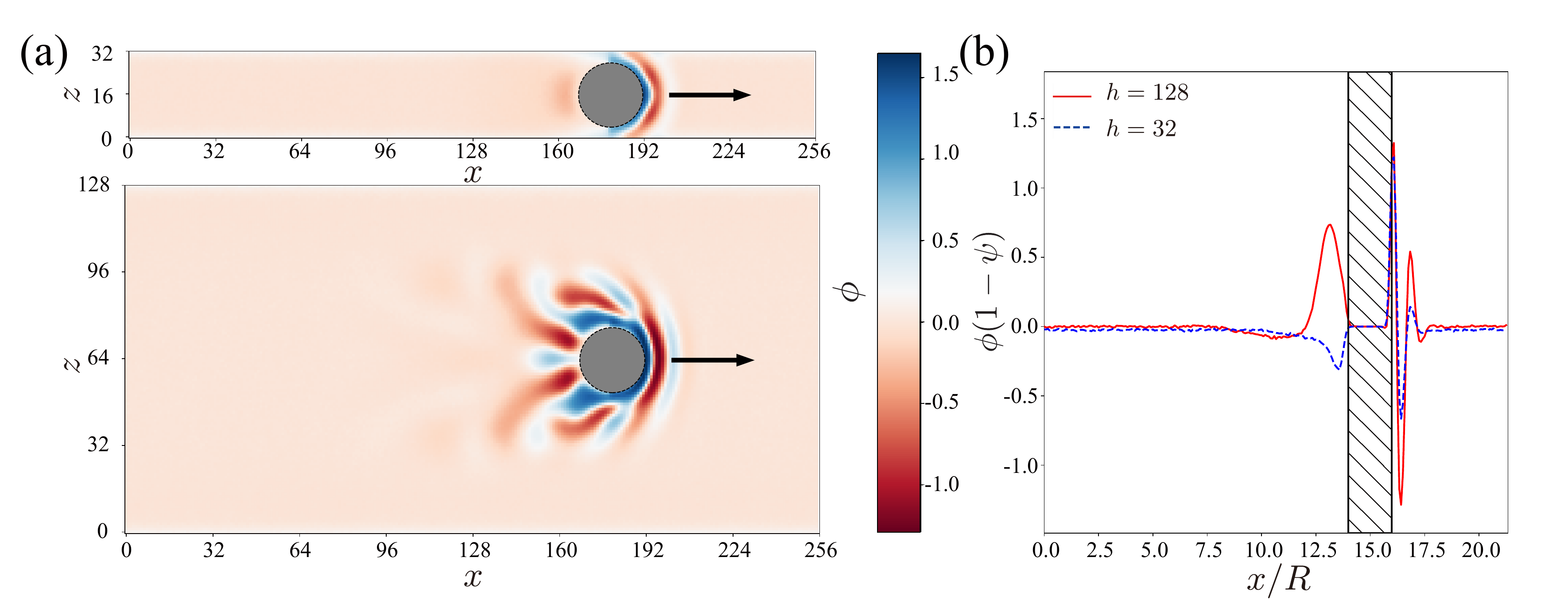}
       \caption{ (a) Snapshots and (b) profiles of the concentration field around the Janus particle corresponding to the temperature field shown in Fig.~\ref{fig:2} at time $t=100$ for two channels of widths $h=32$ (top) and $h=128$ (bottom). The wall temperature is $\tau=-1$ and the heating power  coefficient $g=50$. The average concentration is $\bar{\phi}=0$, and the walls are slightly hydrophilic with $h_{\mathrm{w}}=-0.4$. The black circle represents the particle, and the black arrow shows its orientation. The hatched region in panel (b) indicates the interior of particle.}
\label{fig:3}
\end{figure*}
\begin{figure}[t]
\centering
 \includegraphics[width=0.99\columnwidth]{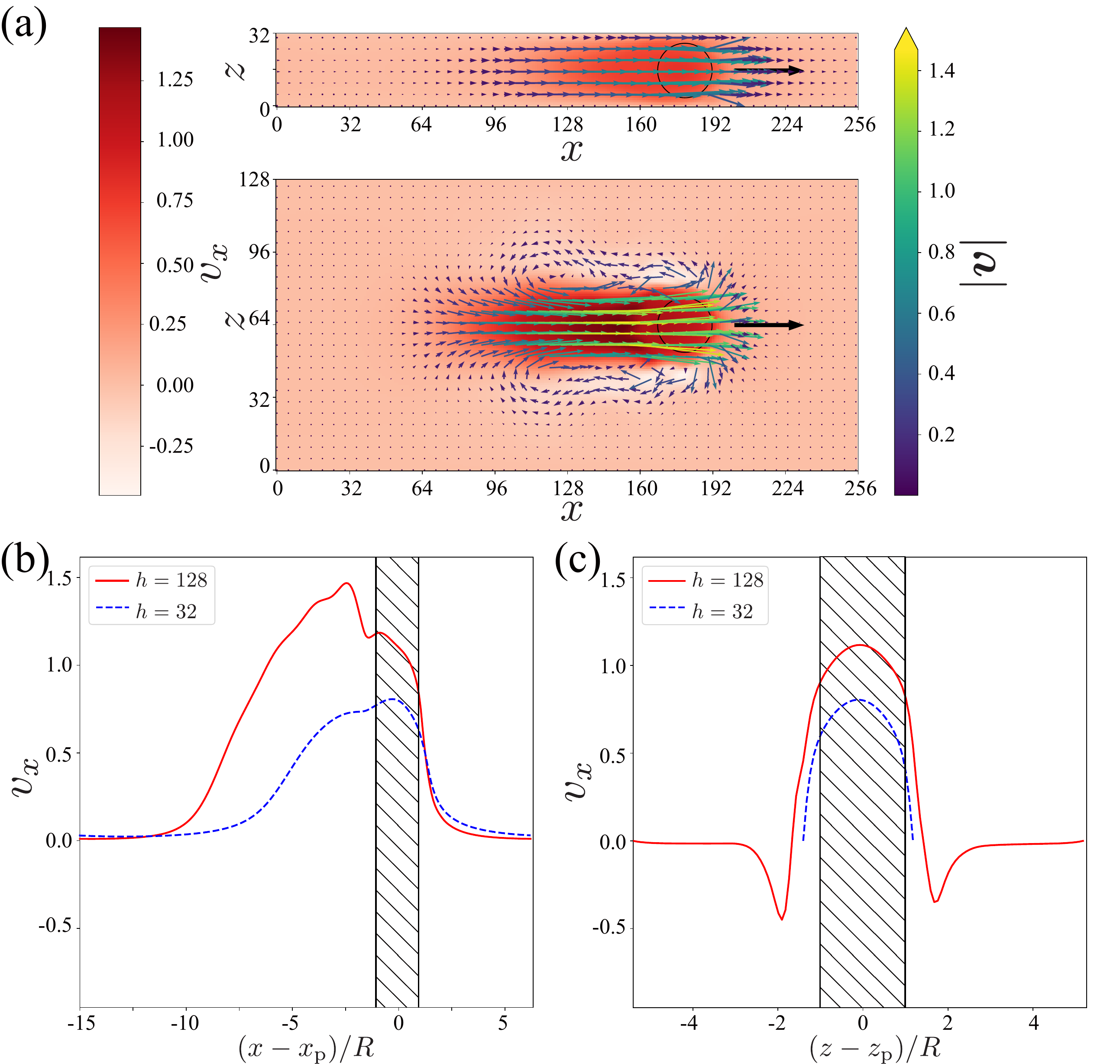}
    \caption{(a) Snapshots of the flow field around the Janus particle corresponding to temperature and concentration fields shown in Figs.~\ref{fig:2} and \ref{fig:3} at time $t=100$  for two channels of widths $h=32$ (top) and $h=128$ (bottom). The arrows represent the local velocity field in the $(x,z)$ plane  and the background color shows the $x$-component of the flow field $v_x$.  The profiles of the $x$-component of the flow velocity $v_x$ along the lines going through the particle center and parallel to the $x$- and  $z$-axis are presented in panels (b) and (c), respectively. The hatched regions indicate the interior of particle. The heating power coefficient $g=50$, and non-slip boundary conditions are imposed for the hydrodynamic flow at the walls. }
    \label{fig:4}
\end{figure}

In general,  a flow of  liquid mixture through the narrow channels, and therefore propulsion of the Janus particle, is influenced by the wetting properties of the channel walls. A component with a greater affinity for the wall tends to accumulate there, which  affects  the concentration distribution within the channel. 
For the chosen weakly hydrophilic walls ($h_\mathrm{w}=-0.4$) the concentration field close to the wall---but far from the Janus particle---is only slightly enhanced. Furthermore, the walls are maintained at a constant temperature $T_\mathrm{i}=0.98\,\Tc$ corresponding to the mixed phase and away from the critical temperature of the solvent.  Therefore, the phenomenon of critical adsorption on the walls is rather  weak here.

Figures~\ref{fig:2} and \ref{fig:3} show snapshots of the temperature and concentration fields at time $t=100$ after switching on the illumination for two channels of widths $h=32$ and $h=128$. In both cases, the initial orientation was parallel to the $x$-axis ($\varphi_{\rm i}=0$) and the heating power coefficient $g=50$.
It is clearly seen that both fields are altered in the narrow channel. Compared to  $h=32$, for $h=128$ the solvent is warmer near the back of the Janus particle (not heated hemisphere), and the region around the particle that is warmer than the surroundings is larger.  For both  channel widths, we see composition layers in front of the moving Janus particle. The first layer is rich in the  component preferred by the heated part of the surface with $\phi>0$. The formation of such concentration layers was observed as a transient phenomenon occurring immediately after irradiation in experiments conducted in the same system but with a Janus particle immobilized by walls \cite{GomezSolano2020a,GomezSolano2020b,Araki2022}. In these experiments, it was observed that these layers disappeared at a later time, and a droplet formed around the heated part of the colloid. Theoretical hydrodynamic description of these experiments revealed strong compositional fluctuations within the droplet \cite{GomezSolano2020a, Araki2022}, a well-known phenomenon in liquids in the presence of stationary temperature gradients~\cite{de2006hydrodynamic}. These anomalously large and long-range thermal fluctuations result from the coupling between temperature and velocity fluctuations. Our results for $g=50$ show that for both channel widths layers always occur in front of the moving Janus particle, and the droplet does not have time to form. Near the uncapped part of the Janus particle, which has no adsorption preferences ($W_{\rm u}=0$), we see both phases (with positive and negative $\phi$). In the case of the wide channel ($h=128$) the concentration field near the uncapped hemisphere  of the particle fluctuates strongly, while in the narrow channel ($h=32$) these fluctuations are barely visible. This is due to the much larger temperature gradient observed in the wide channel. Panel (b) in Fig.~\ref{fig:3} shows the profiles of the concentration field $\phi$ along a straight line parallel to the $x$-axis passing through the center of the particle, corresponding to the snapshots in panel (a).

\begin{figure}[t]
\centering
    \includegraphics[width=0.99\columnwidth]{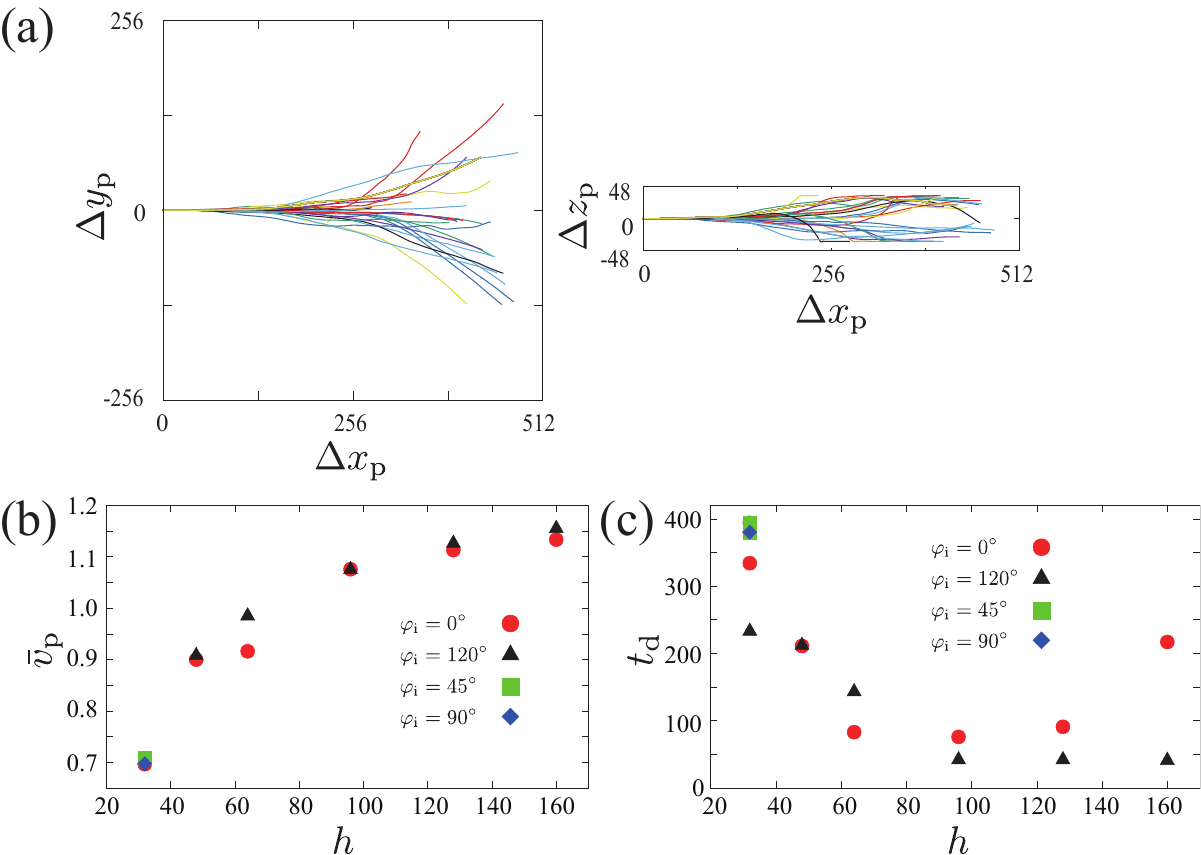}
    \caption{ (a) Typical trajectories of the Janus particle in the channel of the width $h=96$ for the initial angle $\varphi_{\rm i}=0$, recorded up to time $t=500$.  The angle $\varphi$ describes the deviation of the initial orientation of the particle from the $x$-axis in the $xy$-plane, as shown schematically in Fig.~\ref{fig:1}(b). For the sake of clarity, the periodic boundary conditions in $x$-direction has been unraveled. (b) Steady-state velocity of a Janus particle as a function of channel width $h$ for different initial orientations $\bfn_\mathrm{p}$ (different angles $\varphi_{\rm i}$).  (c) Time (in units of $t_0$) needed for the Janus particle to deviate from the straight line motion by at least 1  degree as a function of the channel width.  Heating power  coefficient is $g=50$. }
    \label{fig:5}
\end{figure}

Figure \ref{fig:4} shows snapshots of the flow field around a Janus particle in steady state ($t=100$) for channel widths $h=32$ and $h=128$ corresponding to the temperature and concentration fields shown in Figs.~\ref{fig:2} and \ref{fig:3}. It can be seen that the direction of the motion of particle coincides with its orientation. In narrow channels, the fluid flow is disturbed by the walls, for example, there is no reverse flow around the particle in the $xz$-plane.
Furthermore, in a wider channel, the particle moves faster. For this latter case, the flow fields around spherical particles remains that of a neutral squirmer (source dipole force field)~\cite{Zottl_2016}.
 
In Fig.~\ref{fig:5}(a), typical trajectories of a Janus particle in a channel of width $h=96$ for an initial angle $\varphi_{\rm i}=0$ is presented in the $xy$- and $xz$-planes, whereas in Fig.~\ref{fig:5}(b) we plot the mean particle velocities for  heating power coefficient $g=50$ as a function of the channel width for different values of the initial orientation $\varphi_{\rm i}$ in the $xy$-plane.
First, we see that, generally, the mean velocity is independent of the initial orientation $\varphi_{\mathrm{i}}$ in the $xy$-plane, which is expected. Furthermore, we see that the propulsion velocity decreases as the constraint increases. 

\begin{figure*}[t] 
  \includegraphics[width=0.6\textwidth]{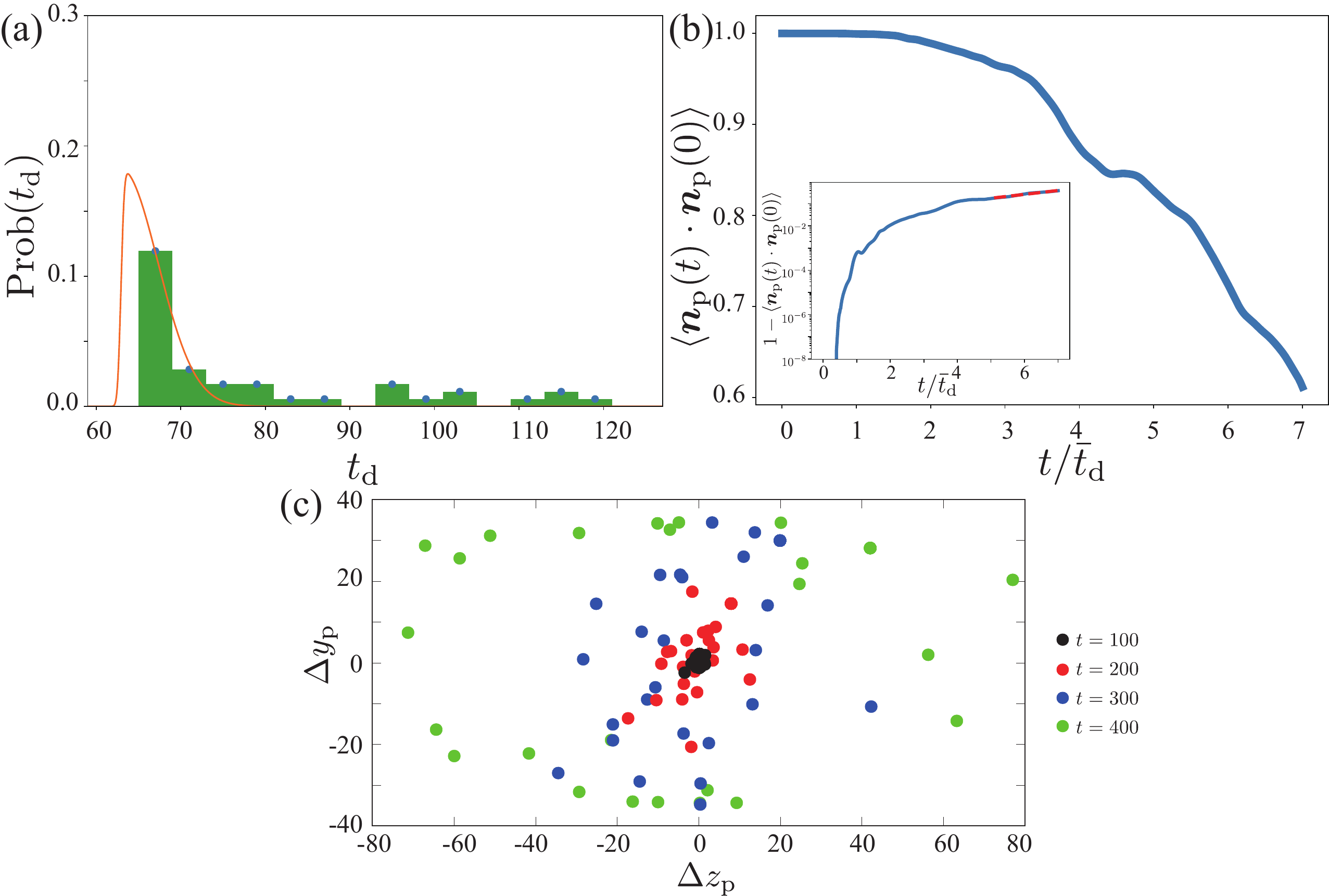}
  \caption{(a)~Histogram of the time $t_{\mathrm{d}}$ elapsed from the moment the laser illumination was switched on until the Janus particle changed direction of motion by at least 1 degree. The line shows the time fit $t_{\mathrm{d}}$ to a log--normal distribution with mean~78.5, standard deviation~17.45, median~69 and skewness~1. (b)~The auto-correlation function of the particle orientation as a function of time rescaled by the mean value of time $t_{\mathrm{d}}$. The inset displays the same data on a linear--logarithmic scale, and the dashed red line corresponds to the linear fit $1-\langle \bm{n}_{\rm p}(t)\cdot\bm{n}_{\rm p}(0)\rangle =\left(5.12\pm 0.04\right)\cdot 10^{-3}(t/\bar{t}_{\rm d})\left(3.74\pm 0.03\right)$. (c)~Distribution of particle displacements in the $yz$-plane at various times. The  width of the channel is $h=96$ and the heating power coefficient $g=50$.}
   \label{fig:6}
\end{figure*}

Another interesting observation is the time it takes for the particle to deviate from its initial direction. The time of first deviation from the original direction of motion of an active particle is primarily determined by its persistence time, which represents the typical time scale on which rotational diffusion or reorientation events significantly alter the direction of motion. We determine this time $t_{\mathrm{d}}$ from a single trajectory and plot it as a function of the channel width $h$ in Fig.~\ref{fig:5}(c) for different values of the initial orientation $\varphi_{\rm i}$ in the $xy$-plane. We observe a trend that the wider the channel, the faster the particle changes its orientation, which is quite surprising.
We see some scatter in the points corresponding to different orientations. This is likely due to the fact that the results were obtained from a single trajectory. (In some cases, this is also due to periodic boundary conditions.)
To investigate the statistical fluctuation further, we determine the distribution for the time elapsed from the switching  heating on until the first deviation of the colloid motion direction from the initial orientation  along the $x$-axis $(\varphi_{\rm i}=0)$ for a channel width $h = 96$ and the heating power coefficient $g = 50$. Figure~\ref{fig:6}(a) shows a histogram obtained from 45 statistically independent trajectories  shown in Fig.~\ref{fig:5}(a) for the time $t_{\rm d}$ needed for the particle to deviate from its path by a predetermined angle (in this case 1 degree). We can see that this is a right-skewed distribution with a long  tail.

In panel~(b) we plot the auto-correlation function of the particle orientation $\left<\bfn_{\mathrm{p}}\left(t\right)\cdot\bfn_{\mathrm{p}}\left(0\right)\right>$ as a function of time rescaled by the mean value $\bar{t}_{\mathrm{d}}$ determined by fitting the histogram to the log--normal distribution (shown by yellow line in panel~(a)). This figure clearly demonstrates that the direction of the trajectory is preserved on the timescale $\bar{t}_{\mathrm{d}}$. For longer times ($t/\bar{t}_{\mathrm{d}} > 5$), the auto-correlation function begins to exhibit an exponentially rapid deviation from the value of 1---see the inset in Fig.~\ref{fig:6}(b), where we use linear--logarithmic scale.
For even longer times, we observe that the particle hits the wall and stays there performing small oscillations. 
\begin{figure}[t]
    \includegraphics[width=0.99\columnwidth]{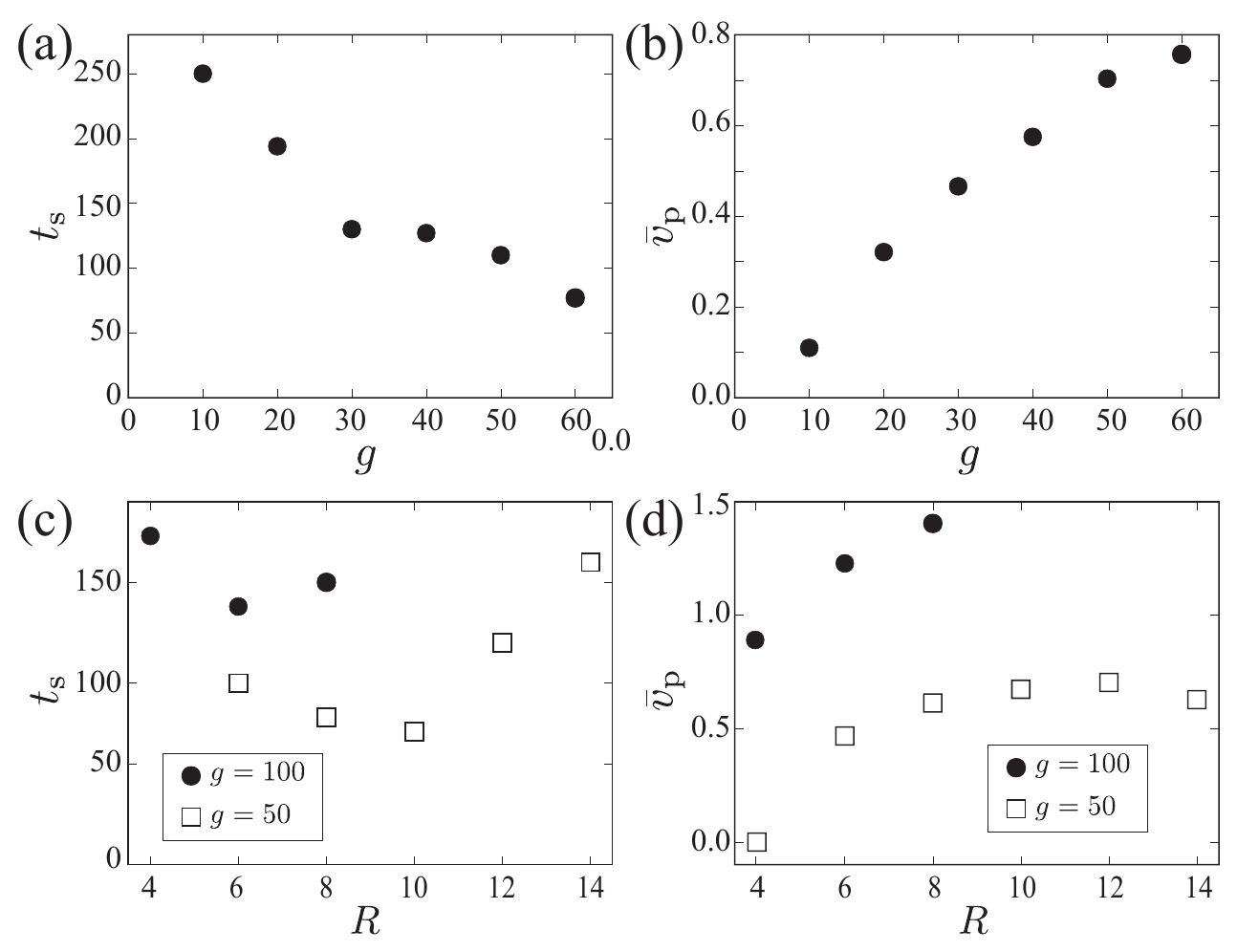}\\

    \caption{ Time $t_{\rm s}$ required to reach the steady state for a particle of radius $R=12$ in a channel of width $h=32$ as a function of (a) heating power $g$ and  (c) the particle radius $R$. Panels (b) and (d)  show the mean speed of the particle $\bar{v}_{\rm p}$ in stationary state for the same narrow channel as a function of heating power $g$ and the particle radius $R$, respectively. 
    }
    \label{fig:h_v1}
\end{figure} 

\subsection{Narrow channel}
\label{subsec:2}

Here and in the next two sections, we focus on the motion of a particle in a narrow channel of width $h=32$.
Unless otherwise stated, all other parameters are the same as in Sec.~\ref{subsec:res1}.

First, we consider a Janus particle with radius $R=12$, initially oriented along the $x$-axis, and analyze the time elapsed from the start of the simulation until the particle reaches the steady-state velocity $\bar{v}_\mathrm{p}$ as a function of the heating power $g$. Figure \ref{fig:h_v1}(a) shows that this time decreases quite rapidly as the parameter $g$ increases. For values of $g > 60$, strong velocity fluctuations are observed, implying that the Janus particle does not, in practice, reach a steady state (see also subsection \ref{subsec:res4}). Figure \ref{fig:h_v1}(b) demonstrates that the steady-state velocity $\bar{v}_{\rm p}$ increases non-linearly with the heating power $g$, although the deviation from linearity is rather small.

The dependence of $t_{\mathrm{s}}$ and $\bar{v}_{\mathrm{p}}$ on the particle radius is more complex. This dependence is illustrated in Fig.~\ref{fig:h_v1}(c)  for heating powers of $g=100$ (squares) and $g=50$ (circles). For the heating power $g=100$, results are presented only for particle radii $R \leqslant 8$; for particles with larger radii at this value of $g$, velocity fluctuations are so large that determining $\bar{v}_{\mathrm{p}}$—and consequently $t_{\mathrm{s}}$—proves impossible. We observe—as is intuitively understandable—that larger particles generally require more time to reach a steady-state velocity. Deviations from this trend may arise from statistical fluctuations. It should be noted that both $t_{\rm s}$ and $\bar{v}_{\rm p}$ were determined based on a single trajectory. The steady-state particle velocity is shown in Fig.~\ref{fig:h_v1}(d). We observe an initial increase in $\bar{v}_{\rm p}$ with particle size, followed by saturation at larger particle sizes.

\begin{figure}[t]
    \centering
     \includegraphics[width=0.99\columnwidth]{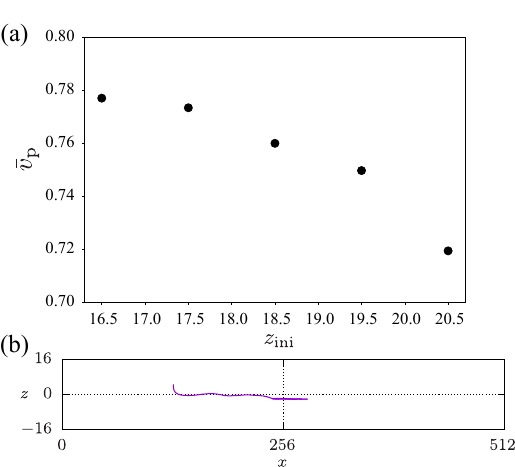}
    \caption{(a) The steady-state velocity $\bar{v}_{\rm p}$ of a particle as a function of its initial position on the $z$-axis. The initial orientation of the particle is aligned with the $x$-axis, and the heating is proportional to $\tau_{\rm s}-\tau$ with an amplitude of $g=100$.
The walls are weakly hydrophobic, with $h_\text{w}=0.4$. (b) Trajectory $z\left(x\right)$ of the Janus particle from $t=0$ to  $t=350$. The particle is initially positioned at $z=20.5$ with the center of the channel located at $z_{\text{center}}=16.5$.}
    \label{fig:z_in}
\end{figure}
\begin{figure*}[t]
    \centering
\includegraphics[width=0.9\textwidth]{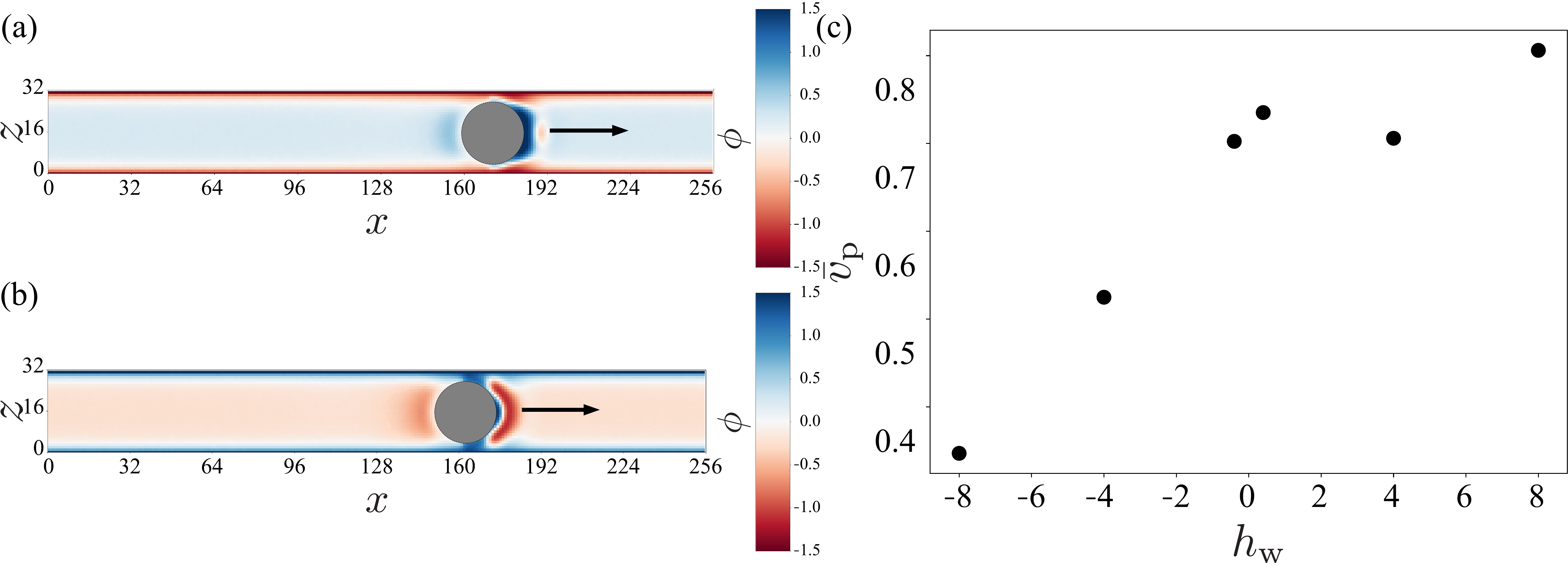}
\caption{  Snapshots  of the concentration field around the Janus particle for the case of  (a)~strongly hydrophobic walls ($h_\mathrm{w}=8$) and (b)~strongly hydrophilic walls ($h_\mathrm{w}=-8$). The snapshots were taken at $t=100$ for the channel of width $h=32$. (c)~The steady-state velocity of the particle increases with the increasing parameter $h_\mathrm{w}$. The initial orientation of the particle is aligned with the $x$-axis, and the heating parameter $g=50$. The particles were initially positioned at the center of the channel. }
    \label{fig:hw}
\end{figure*}

We find that when the particle's initial position is shifted toward one of the walls (along the $z$-axis), the colloidal particle experiences a repulsive force under illumination that drives it back to the center of the channel. We also observe that such an initial displacement slightly reduces the steady-state velocity, as illustrated in Fig.~\ref{fig:z_in}(a). This effect may stem from small oscillations of the particle in the $z$-direction, visible in Fig.~\ref{fig:z_in}(b).

\subsection{Effects of adsorption films on the walls}
\label{subsec:res3}

In the previous sections, we considered the wetting parameter $h_{\rm w}$---defining the preference of the wall for one of the solvent components---which was very weak compared to the wetting parameter $W_{\rm c}$ of the capped hemisphere of the Janus particle.

In the present section, we increased the value of the channel wall wetting parameter, considering cases of $h_{\mathrm w}=\pm 4$ and $\pm 8$. For $h_{\mathrm w}= \pm 8$, the adsorption strength matches that of the capped part of the particle ($W_\mathrm{c}=-8$). This enhanced adsorption preference leads to the formation of thicker layers enriched in the preferred phase on the walls. We anticipate that, in a narrow channel, this could significantly affect the phase separation structure around the particle and, consequently, its self-propulsion speed. This is indeed the case, as illustrated in Fig.~\ref{fig:hw}. The observed concentration distribution depends crucially on whether the walls exhibit a preference for the same phase of the binary solvent as the capped hemisphere of the colloid, or for the opposite phase. If the walls prefer the same phase, a region of that phase formed around the heated part of the colloid connects to the adsorption layers---see Fig.~\ref{fig:hw}, panel~(b). This causes the particle to slow down. In the opposite scenario, shown in panel~(a), the particle moves slightly faster (see graph in Fig.~\ref{fig:hw}(c)).

\subsection{Viscosity effects}
\label{subsec:res4}
In this section, we investigate the effect of solvent viscosity on particle dynamics. In the previous sections, we focused on the relatively low-viscosity case of $\eta = 50$. For a particle driven by an external force, the particle velocity is generally expected to increase as the solvent viscosity decreases. In the present system, however, the situation is more complicated because the hydrodynamic flow is strongly coupled to the particle motion, as well as to the concentration and temperature fields. Therefore, the effect of solvent viscosity on the particle dynamics is not straightforward. 

\begin{figure}[t] 
\centering
   \includegraphics[width=0.99\columnwidth]{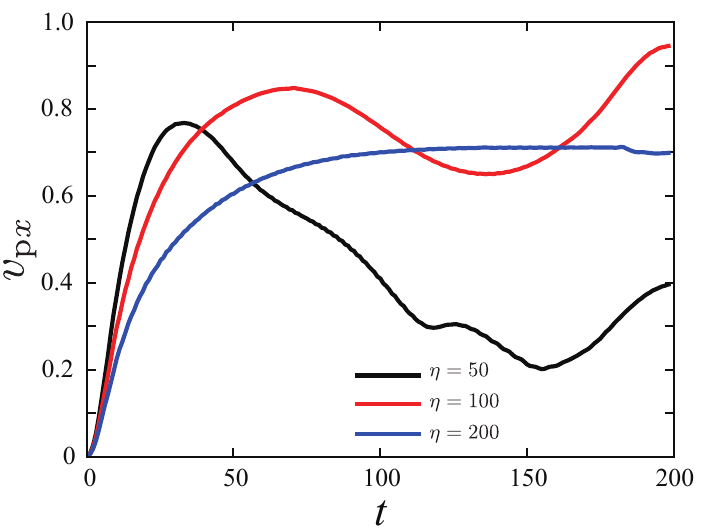}\\
   \caption{Temporal changes of the particle velocity along its initial orientation in binary solvents with viscosities $\eta = 50$, $100$, and $200$. A particle with radius $R = 12$ is placed in a channel of width $h = 32$.
   The heating power coefficient is $g=100$. }
   \label{fig:8}
   \end{figure}

Figure \ref{fig:8} shows the temporal evolution of the particle velocity toward the initial direction of the particle for different solvent viscosities $\eta$. 
Here, we fix the ratio $\eta_{\rm p}/\eta=50$. Initially, a particle with radius $R=12$ is placed at the midplane of a channel with width $h=32$. The lateral system size is $256^2$, the heating power is $g=100$, and the other parameters are identical to those used in Sec.~\ref{subsec:res1}.

In the early stage, after switching on the illumination, the particle speed increases for all viscosities, and the acceleration is larger for lower viscosities. However, in the low-viscosity cases, the particle speed subsequently exhibits large fluctuations. In contrast, for higher viscosities, the increase in velocity is more gradual, and the system eventually reaches a steady state. As a result, the net particle speed in the high-viscosity case can become comparable to that in the low-viscosity case. 
The particle speed is not simply proportional to the inverse of the solvent viscosity.

\begin{figure*}[t] 
\centering
   \includegraphics[width=0.8\textwidth]{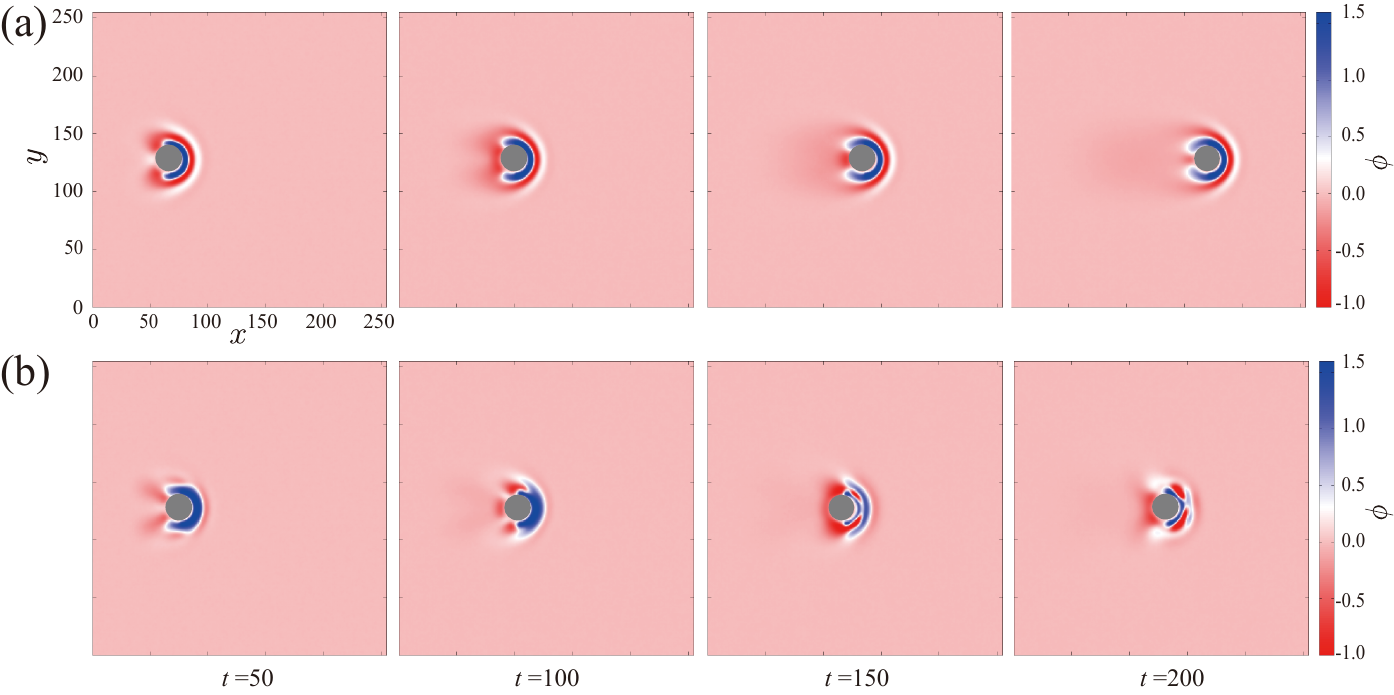}\\
      \caption{The development of the concentration field $\phi$ in the $xy$-plane in the binary solvents of the viscosity (a) $\eta=200$ and (b) $\eta=50$. }
   \label{fig:9}
   \end{figure*}

Figure \ref{fig:9} shows typical concentration patterns in the $xy$-plane for the cases of higher and lower viscosity.
For $\eta = 200$, in panel (a), a well-defined layered structure forms near the head side of the particle at an early stage. The system subsequently reaches a steady state in which the particle moves continuously along the head direction. 
In contrast, for $\eta = 50$, in panel (b), the concentration layers initially develop but later exhibit strong fluctuations. The layered structure is transiently disrupted and repeatedly reconstructed. In the low-viscosity case, hydrodynamic flows around the particle become stronger, and these flows likely destabilize the concentration layers. As a result, the repeated reorganization of the concentration pattern disturbs the particle motion, leading to the large velocity fluctuations visible in Fig.~\ref{fig:8}.

\begin{figure*}[t] 
\centering
   \includegraphics[width=0.8\textwidth]{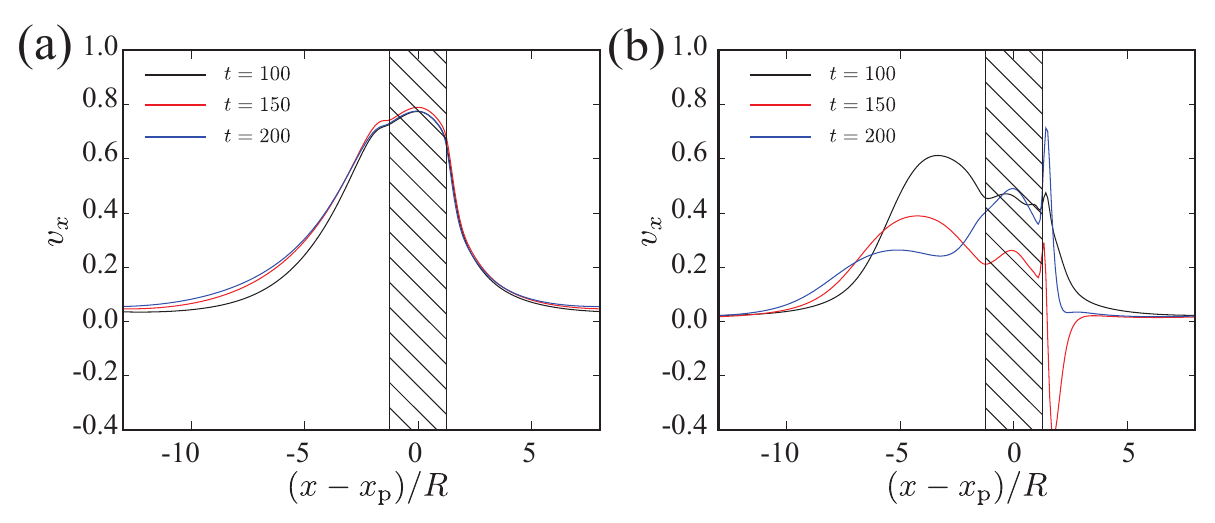}
   \caption{The evolution of the velocity profile along the motion in the particle frame. 
   The solvent viscosity is set to (a) $\eta=200$ and (b) $\eta=50$. 
   The hatched region indicates the interior of particle.}
   \label{fig:10}
   \end{figure*}

The hydrodynamic wake generated by the particle motion may disturb the motion of surrounding particles. 
Figure \ref{fig:10} shows the temporal evolution of the velocity-field profile along the direction of motion in the particle frame. 
As shown in Fig.~\ref{fig:9}, the concentration field becomes steady in the high-viscosity case, and the resulting velocity field, presented in Fig.~\ref{fig:10}(a), is also stationary. 
In addition, the velocity field behind the moving particle decays relatively quickly, although it is still slightly affected by the periodic boundary condition.
On the other hand, in the low-viscosity case shown in Fig.~\ref{fig:10}(b), the velocity field exhibits significant temporal fluctuations, leading to the fluctuating particle motion shown in Fig.~\ref{fig:8}. 
A non-negligible hydrodynamic wake remains behind the moving particle, and it tends to grow with time.  Hydrodynamic flows dissipate with the kinematic viscosity $\eta/\rho$, therefore, in the low-viscosity case, the velocity field is not fully dissipated into the background solvent, and the remaining hydrodynamic flow can potentially disturb the motion of particles surrounding the focused particle.

\section{Discussion and conclusions}
\label{sec:dis}

In this section, we will discuss our results in relation to the experimental observations.

The studies~\cite{Volpe2011,Buttinoni_2012,gomez2017tuning} investigated various aspects of the motion of light-activated Janus spheres within a critical binary mixture. The system was placed between two microscope slides, a configuration that geometrically corresponds to a channel. Among other parameters, the mean velocity $v_{\mathrm{p}}$ and the average timescale $t_{\text{dir}}$---during which the direction of trajectory is maintained---were determined as functions of illumination intensity and particle size. The authors determined these quantities by fitting the measured mean squared displacements (MSD) to the theoretical predictions of the ABP model (in a two-dimensional (2D) projection): $\delta r_{\mathrm{p}}^2=\left(4D_{\mathrm{p}}+\ell^2/t_{\text{dir}}\right)\delta t + \ell^2/2\,\left[\exp(-2\delta t/t_{\text{dir}})-1)\right]$, where $\ell$ denotes the distance traveled during swimming, $v_{\mathrm{p}}=\ell/t_{\text{dir}}$, and $D_{\mathrm{p}}$ is the particle diffusion constant. In general, this prediction should characterize the motion of self-propelled particles, which on short time scales ($\delta t\ll t_{\text{dir}}$) exhibits ballistic behavior, with $\delta r_{\mathrm{p}}^2\simeq v_{\mathrm{p}}^2\delta t^2$, while on long time scales ($\delta t \gg t_{\text{dir}}$) it becomes effectively diffusive, with $D_{\text{eff}}=D_{\mathrm{p}}+4\ell^2/t_{\text{dir}}$~\cite{PhysRevLett.99.048102}.

At this point, we wish to emphasize that the trajectories in the experiments were recorded over a period of the order of \qty{1000}{\second} or longer, which is not feasible within a reasonable time frame in the context of our simulations. 
The typical duration of our simulations was $1000\,t_0$, where $t_0 = d_0^2/L_0$. Converting the dimensionless quantities we adopted in our simulations---described in Sec.~\ref{sec:2} and dictated by computational limitations---into physical quantities yields very short time and length scales.
Therefore, we can only investigate the short-time behavior of an active Janus particle, dominated by ballistic motion due to propulsion and the crossover to the enhanced diffusion at longer times as can be seen in Fig.~\ref{fig:5}(a). Consequently, we cannot determine the average timescale $t_{\rm dir}$ via a fitting procedure---as is done in experiments---but we can determine the time $t_{\rm d}$ at which the first deviation from rectilinear motion occurs.  On the other hand, the phenomena focused on here arise primarily from the relative importance of advection compared to diffusion; these quantities are characterized by the P\'eclet numbers for composition and temperature---${\operatorname{Pe}}_{\phi}=Rv_{\mathrm{p}}/L_0 \simeq 1$ and ${\operatorname{Pe}}_{\mathrm{T}}=Rv_{\mathrm{p}}/L_{\mathrm{T}0}$ (with ${\operatorname{Pe}}_{\phi}/{\operatorname{Pe}}_{\mathrm{T}}\gg 1$), respectively---which is consistent with experimental values~\cite{Volpe2011,gomez2017tuning}. Moreover, an advantage of our simulations is that we can record trajectories in three dimensions, whereas digital video microscopy, used in experiments, only tracks the projection of particle trajectories onto a two-dimensional plane.  

We observe that the Janus-type particle changes its direction of motion in all three dimensions, occasionally approaching one of the walls. Its orientation remains virtually unchanged  until the first deviation from the initial direction at time $t_{\rm d}$---see Fig.~\ref{fig:6}(b). However, the very first deviation from the initial direction of motion occurs isotopically in all directions---see Fig.~\ref{fig:6}(c).
This differs from the phenomenon of ``orientational quenching'' observed for other types of active particles~\cite{das2015boundaries} and also hypothesized in~\cite{gomez2017tuning}.
Ultimately, typically toward the end of our observation period or longer (depending on the initial orientation in the $xy$-plane) the Janus particle reaches one of the walls and executes small movements along it. Under stronger illumination, fluctuations in the concentration and velocity fields are more pronounced, and the particle reaches the wall sooner.

In experiments reported in Ref.~\cite{Volpe2011},  the authors observed that the average time scale $t_{\rm dir}$ over which the trajectory direction is maintained is close (within $\pm 10\%$) to the timescale of rotational diffusion $t_{\rm rot}$ for an unilluminated particle, \ie, when it undergoes normal Brownian motion. For example, for a particle radius $R=\qty{2.13}{\micro\meter}$ and a cell width to particle radius ratio $h/R=7/2.13=3.2(9)$, the authors obtained $t_{\text{rot}}/t_{\text{dir}}\simeq 0.9$. Similar agreement between these timescales was also observed for particles with radii $R=\qty{1.0}{\micro\meter}$ and $R=\qty{0.5}{\micro\meter}$, corresponding to ratios of $h/R=7$ and $14$, suggesting that the channel width does not affect rotational diffusion.
Interestingly, no significant changes in the mean timescale $t_{\mathrm{dir}}$ were observed as the illumination intensity varied.
(It is worth noting that the time $t_{\mathrm{rot}}$ was determined based on the relationship $1/t_{\text{rot}}=D_{\text{rot}} =3D_{\mathrm{p}}/\left(4R^2\right)$, which links the rotational and translational diffusion coefficients for a sphere, and the measured value of $D_{\mathrm{p}}$. The authors of Ref.~\cite{Volpe2011} reported only the value $D_{\mathrm{p}}=\qty{0.031\pm 0.006}{\micro\meter\squared\per\second}$, measured for a system with a ratio of $h/R=7/2.13=3.2(9)$; this value is approximately 50\% lower than the corresponding values derived from the Stokes--Einstein equation, a discrepancy attributed to hydrodynamic coupling with the walls~\cite{happel1973low}.) Similar conclusions regarding the ratio $t_{\text{rot}}/t_{\text{dir}}$  were drawn in~\cite{gomez2017tuning} for particles of various sizes, studied at a constant ratio of $h/R=4$.

In our simulations, we are unable to determine $D_{\text{rot}}$ by fitting to the MSD formula. However, we observe that the mean time of the first deviation from rectilinear motion $t_{\mathrm{d}}$ depends distinctly on the channel width. For the parameters considered in Sec.~\ref{subsec:res1}, in the narrowest channel ($h=32$), this time is several times longer than in a channel of width $h=128$. This demonstrates that, in the case of this type of active colloid, geometric confinement promotes directed motion by inhibiting rotational diffusion. 

Finally, we also investigated the influence of solvent viscosity on particle motion. Although it is commonly expected that low-viscosity fluids would enhance the efficiency of particle motion control, the situation is not quite so straightforward. In low-viscosity fluids, the concentration field is readily modified by hydrodynamic flows, resulting in significant fluctuations in the induced particle velocity. Furthermore, due to insufficient dissipation of the hydrodynamic flow, a strong hydrodynamic wake persists behind a moving particle. These effects influence the motion of surrounding particles and compromise the ability to control the movement of the target particle.

Our studies demonstrate that the FPD method can contribute to a physical understanding of the dynamic behavior of Janus colloids undergoing self-propulsion in a confined binary solvent. The application of this method will enable the verification of certain predictions derived from a thermodynamic approach to the selfdiffusiophoresis of colloidal Janus particles~\cite{PhysRevE.99.060602}. Since this method can be readily extended to the multi-particle case, we intend to utilize it in the future to investigate interactions between pairs of active colloids, and subsequently to analyze their collective behavior. This will make it possible to validate numerous existing results obtained within the framework of simplified models of active particles.

\section{Data Availability Statement}
The data that supports the findings of this study are available from the corresponding author upon reasonable request.

\begin{acknowledgments}
The research of M.P. and A.M. was funded by  the  National Science Center, Poland (Opus Grant  No.~2022/45/B/ST3/00936). 
T.A. was supported by JST CREST Grant Number JPMJCR2095, JSPS KAKENHI 25K00967 and 24K00592, and The
University of Tokyo ISSP for the provision of computer time.
\end{acknowledgments}

\appendix

\renewcommand{\thefigure}{\Alph{section}.\arabic{figure}}
\setcounter{figure}{0}

\section{Constant heating power}
\label{app:A}

\begin{figure}[b]
         \includegraphics[width=0.99\columnwidth]{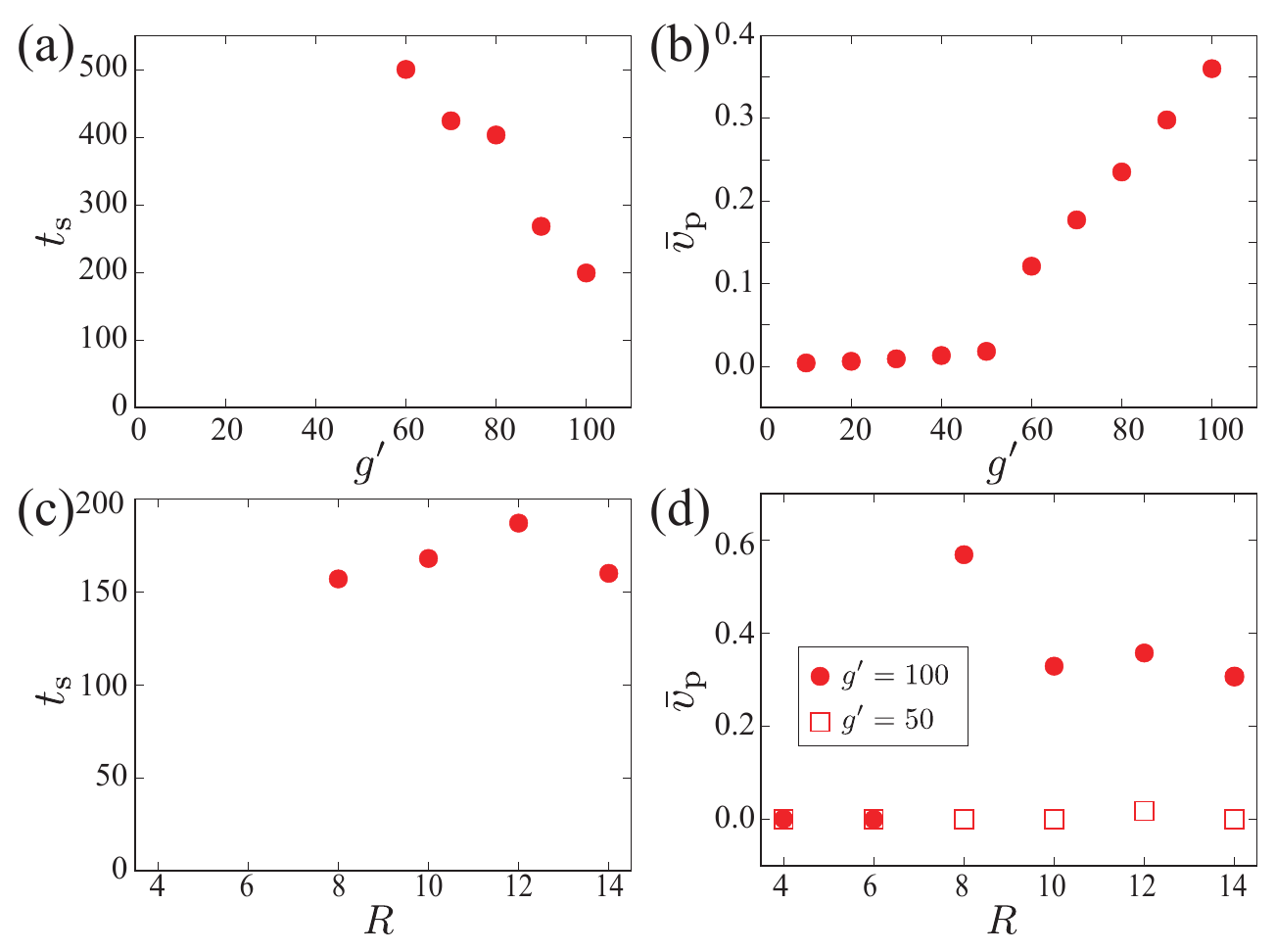}

    \caption{(a) Time $t_{\rm s}$ required to reach the steady state for a particle of radius $R=12$ in a channel of width $h=32$ as a function of heating power $g'$ (a) and the particle radius $R$ (c). The mean speed of the particle $\bar{v}_{\rm p}$ in stationary state as a function of heating power $g'$  (b) and  the particle radius $R$ (d).
 The heating rate is independent of surface temperature. 
    }
    \label{fig:h_v}
\end{figure} 

\begin{table*}[t]
    \caption{Values of parameters used in the simulations. We use $d_0$, $t_0=d_0^2/L_0$, and $\kB\Tc$ as units of distance, time, and energy, respectively. Wherever possible, we additionally give reference to equation where the quantity appears for the first time.}
    \label{tbl:parameters}
\begin{ruledtabular}
\begin{tabular}{lcccc}
Quantity & Symbol & Numerical value & Unit & Equation \\
\hline
Size of system in $x$ and $y$ direction      & $N_x$, $N_y$ & $256$ & $d_0$ &\\
Size of system in $z$ direction              & $h$ & \numrange{32}{128} & $d_0$ &\\
Time step                                    & $\Delta t$ & $0.001$ & $t_0$ &\\
Total time of simulation                     & $t_\text{max}$ & $1000$ & $t_0$ &\\
Temperature on walls                         & $T_\mathrm{i}$ & $0.98$ & $\Tc$ &\\
Radius of particle                           & $R$ & \numrange{4}{14} & $d_0$ & \\
Width of the surface region                  & $d_\psi$ & $2$ & $d_0$ & \eqref{eq:particle} \\
Angular width of orientation function        & $d_\theta$ & $0.033$ & 1 & \eqref{eq:orientation} \\
Coefficient of gradient of order parameter   & $c$ & $4$ & $d_0^2\kB\Tc$  & \eqref{eq:e_phi} \\ 
\multirow{2}{*}{Solvent invasion parameters} & $\chi_\mathrm{p}$ & $5$ & $\kB\Tc$ & \eqref{eq:e_phi} \\ 
                                             & $\phi_\mathrm{p}$ & $0$ & $1$ & \eqref{eq:e_phi} \\ 
Wetting parameter of uncapped hemisphere     & $W_\mathrm{u}$ & $0$ & $\kB\Tc$ & \eqref{eq:e_phi} \\ 
Wetting parameter of capped hemisphere       & $W_\mathrm{c}$ & $-8$ & $\kB\Tc$ & \eqref{eq:e_phi} \\
$\phi^4$ entropy parameter                   & $b$ & 1 & 1 & \eqref{eq:entropy} \\
Strength of noise                            & $\zeta_0$ & $10^{-4}$ & $d_0/t_0$ & \eqref{eq:dev_phi} \\
Thermal diffusion constant for solvent       & $L_{\mathrm{T}0}$ & $100$ & $L_0$ & \eqref{eq:dev_T}\\
Thermal diffusion constant for particle      & $L_{\mathrm{Tp}}$ & $300$ & $L_0$ & \eqref{eq:dev_T}\\
Mass density  & $\rho$ & 0.01 & $k_{\rm B}T_{\rm c}/d_0L_0^2$ & \eqref{eq:dev_v} \\
Heating power coefficient                    & $g,g'$ & \numrange{10}{100} & $1/t_0$ & \eqref{eq:cool_pow}\\
Target temperature of capped hemisphere      & $\tau_\mathrm{s}$ & 6.631 & 1 & \eqref{eq:cool_pow}\\
Solvent viscosity                            & $\eta$ &  50, 100, 200 & $\rho L_0$ & \eqref{eq:vstress}\\ 
Ratio of particle and solvent viscosity                         & $\eta_\mathrm{p}/\eta$ & 50 & 1 & \eqref{eq:vstress}\\ 
Strength of wall--particle interaction       & $\epsilon_\mathrm{w}$ & 0.25 & $\kB\Tc$ & \eqref{eq:wall_potential}\\
Wetting parameter of wall                    & $h_\mathrm{w}$ & \numrange{-8}{8} & $\kB\Tc$ & \\
\end{tabular}
\end{ruledtabular}
\end{table*}

\begin{table*}[t]
    \caption{Values of the parameters that were varied in our simulations.}
    \label{tbl:parameters2}
\begin{ruledtabular}
\begin{tabular}{lccccc}
Figure & $h\,\left[d_0\right]$ & $R\,\left[d_0\right]$ & $g,g'\,\left[1/t_0\right]$ & $\eta\,\left[\rho L_0\right]$ & $h_\mathrm{w}\,\left[\kB\Tc\right]$ \\
\hline
Figs.~\ref{fig:2}, \ref{fig:3}, and \ref{fig:4} & $32$ and $128$ & $12$ & $50$ & $50$ & $-0.4$\\ 
Fig.~\ref{fig:5}(a) and \ref{fig:6} & $96$ & $12$ & $50$ & $50$ & $-0.4$ \\
Fig.~\ref{fig:5}(b) and (c) & from $32$ to $160$ & $12$ & $50$ & $50$ & $-0.4$ \\
Fig.~\ref{fig:h_v1}(a) and (b) & $32$ & $12$ & from $10$ to $100$ & $50$ & $-0.4$ \\
Figs.~\ref{fig:h_v1}(c) and (d) & $32$ & from $4$ to $14$ & $50$ and $100$ & $50$ & $-0.4$\\
Fig.~\ref{fig:z_in} & $32$ & $12$ & $100$ & $50$ & $0.4$ \\
Fig.~\ref{fig:hw}(a) and (b) & $32$ & $12$ & $50$ & $50$ & $8$ and $-8$\\
Fig.~\ref{fig:hw}(c) & $32$ & $12$ & $50$ & $50$ & from $-8$ to $8$\\
Fig.~\ref{fig:8} & $32$ & $12$ & $100$ & $50$, $100$, and $200$ & $-0.4$\\
Figs.~\ref{fig:9} and \ref{fig:10} & $32$ & $12$ & $100$ & $200$ and $50$ & $-0.4$ \\
Figs.~\ref{fig:h_v}(a) and (b) & $32$ & $12$ & from $10$ to $100$ & $50$ & $-0.4$ \\
Figs.~\ref{fig:h_v}(c) and (d) & $32$ & from $4$ to $14$ & $50$ and $100$ & $50$ & $-0.4$ \\
\end{tabular}
\end{ruledtabular}
\end{table*}

Here we consider Janus particles in a narrow channel of width $h=32$, whose initial orientation is along the $x$-axis, and analyze their motion for  heating model in Eq.~\eqref{eq:dev_T} with
\begin{equation}
     \label{heat:without} H=g',   
\end{equation}
where the heating parameter $g'$ is constant. We want to compare how the results presented in Sec.~\ref{subsec:2} change if we neglect heat dissipation at the capped hemisphere of the Janus particle.

The time elapsed from the start of the simulation until the particle reaches the steady-state velocity $\bar{v}_\mathrm{p}$ is shown in Fig. \ref{fig:h_v}(a). Compared to the heating model based on the difference $\tau_{\rm s}-\tau(\bfr)$, the time required to reach the steady state is significantly longer.

Figure \ref{fig:h_v}(b) shows the steady-state velocity $\bar{v}_{\rm p}$ as a function of heating power. It can be observed that the particle velocity is close to zero for $g$ values up to 50, whereas above this value, it increases linearly with $g$. This differs from the situation where $H$ is proportional to $\tau_{\rm s}-\tau(\bfr)$; in that case, the increase in $\bar{v}_{\rm p}$ with $g$ deviates from a linear relationship. Furthermore, in the case under discussion, the maximum temperature—and consequently the temperature gradients—are lower than when heat is additionally removed from the particle surface, i.e., when $H$ is defined by equation \eqref{eq:cool_pow} with parameter $g=g'$. As a result, concentration gradients are also smaller, and the Janus particle moves significantly more slowly (compare with  Fig. \ref{fig:h_v1}(b)). Although the velocity "tail" trails the moving particle for a similar distance, the flow dissipates before the particle returns to the same position under periodic boundary conditions.

The dependence of $t_{\mathrm{s}}$ and $\bar{v}_{\mathrm{p}}$ on the particle radius is shown in Figure \ref{fig:h_v}(c) and (d), respectively, for heating powers of $g'=100$ (circles) and $g'=50$ (squares). It can be seen that for $g'=50$, the values of $t_{\mathrm{s}}$ and $\bar{v}_{\mathrm{p}}$ are zero, as the heating power proves insufficient to set the particle in motion—even for particles with larger radii.
For $g'=100$, a trend similar to that observed when $H$ depends on the difference $\tau_{\mathrm{s}}-\tau$ is seen: the value of $\bar{v}_{\mathrm{p}}$ initially increases with particle size but reaches saturation for larger radii $R$.

\section{Parameters used in numerical simulations}\label{app:B}

The values used in numerical simulations for all the parameters of our model are listed in Table~\ref{tbl:parameters}. In Table~\ref{tbl:parameters2} we list the values of the parameters that were varied in our simulations.

\end{document}